%% file: new.tex
\documentclass{aa}
\usepackage{txfonts}
\usepackage{graphicx}
\usepackage{setspace}
\usepackage{natbib}
\usepackage{sidecap}
\bibpunct{(}{)}{;}{a}{}{,}
\usepackage{color}
\usepackage{subfigure}
\usepackage{floatflt}
\usepackage{rotating}
\usepackage{aalongtable, lscape, afterpage}

\newcommand{\rul}{\rule[-2.50mm]{0mm}{9mm}}

\begin{document}

\input{makros}

\title{AGN heating and ICM cooling in the $\hiflux$ sample of galaxy
  clusters}

\author{Rupal~Mittal\inst{1, 2} \and Daniel~S.~Hudson \inst{1} \and
  Thomas~H.~Reiprich\inst{1} \and Tracy~Clarke\inst{3, 4}} \institute{
  Argelander-Institut f\"ur Astronomie, Auf dem H\"ugel 71, 53121
  Bonn, Germany \and Max-Planck-Institute f\"ur Radioastronomie, Auf
  dem H\"ugel 69, 53121 Bonn, Germany \and Naval Research Laboratory,
  Code 7213, 4555 Overlook Ave. SW, Washington, D.~C. 20375, USA \and
  Interferometrics Inc., 13454 Sunrise Valley Drive, Suite 240,
  Herndon, VA 20171 USA}

\date{Received/Accepted}

\abstract{Active galactic nuclei~(AGN) at the center of galaxy
  clusters with gas cooling times that are much shorter than the
  Hubble time have emerged as heating agents powerful enough to
  prevent further cooling of the intracluster medium~(ICM). We carried
  out an intensive study of the AGN heating$-$ICM cooling network by
  comparing various cluster parameters to the integrated radio
  luminosity of the central AGN, $\lr$, defined as the total
  synchrotron power between 10~MHz and 15~GHz. This study is based on
  the $\hiflux$ sample comprising the 64 X-ray brightest galaxy
  clusters. We adopted the central cooling time, $\ct$, as the
  diagnostic to ascertain cooling properties of the $\hiflux$ sample
  and classify clusters with $\ct < 1$~Gyr as strong cool-core~(SCC)
  clusters, with $1$~Gyr~$< \ct <7.7$~Gyr as weak cool-core~(WCC)
  clusters and with $\ct > 7.7$~Gyr as non-cool-core~(NCC) clusters.
  We find 48 out of 64 clusters (75\%) contain cluster center radio
  sources~(CCRS) cospatial with or within 50$\hinv{-1}$kpc of the
  X-ray peak emission. Furthermore, we find that the probability of
  finding a CCRS increases from 45\% to 67\% to 100\% for NCC, WCC,
  and SCC clusters, respectively.
  
  \hspace*{0.5cm} We use a total of $\sim140$ independent radio
  flux-density measurements, with data at more than two frequencies
  for more than 54\% of the sources extending below 500~MHz, enabling
  the determination of accurate estimates of $\lr$. We find that $\lr$
  in SCC clusters depends strongly on the cluster scale such that more
  massive clusters harbor more powerful radio AGN. The same trend is
  observed between $\lr$ and the classical mass deposition rate,
  $\mdr$ in SCC and partly also in WCC clusters, and can be quantified
  as $\lr \propto \mdr^{1.69\pm0.25}$. We also perform correlations of
  the luminosity for the brightest cluster galaxy, $\lbcg$, close to
  the X-ray peak in all 64 clusters with $\lr$ and cluster parameters,
  such as the virial mass, $\mvir$, and the bolometric X-ray
  luminosity, $\lx$. To this end, we use the 2MASS $K$-band magnitudes
  and invoke the near-infrared bulge luminosity-black hole mass
  relation to convert $\lbcg$ to supermassive black hole mass,
  $\mbh$. We find a weak correlation between $\mbh$ and $\lr$ for SCC
  clusters, $\lr \sim \mbh^{4.10 \pm 0.42}$, although with a few
  outliers. We find an excellent correlation of $\lbcg$ with $\mvir$
  and $\lx$ for the entire sample, the SCC clusters showing a tighter
  trend in both the cases. We discuss the plausible reasons behind
  these scaling relations in the context of cooling flows and AGN
  feedback.

  \hspace*{0.5cm} Our results strongly suggest an AGN-feedback
  machinery in SCC clusters, which regulates the cooling in the
  central regions. Since the dispersion in these correlations, such as
  that between $\lr$ and $\mdr$ or $\lr$ and $\mbh$, increases in
  going from SCC to WCC clusters, we conclude there must be secondary
  processes that work either in conjunction with the AGN heating or
  independently to counteract the radiative losses in WCC clusters.

}

\authorrunning{Rupal Mittal et~al.}  \titlerunning{AGN heating in the
  $\hiflux$ sample of Galaxy Clusters}
\maketitle

\section{Introduction}
\label{intro}

In recent years, heating by active galactic nuclei~(AGN) through
outflows has gained fundamental importance in the realm of large-scale
structure and galaxy formation. Several studies
\citep[e.g.][]{Sijacki2007,Croton2006,Bower2006,Scannapieco2005,Silk1998}
have contributed to this comprehensive picture, wherein AGN feedback
is considered an attractive solution to several connected problems,
such as the high-mass end truncation of galaxy distribution
\citep[e.g.][]{Benson2003} and the absence of cooling-flows in centers
of galaxy clusters~\citep[e.g.][]{McNamara2007}. According to these
studies, AGN heating at the centers of clusters may likely be
responsible for quenching condensation of the hot intracluster medium
onto the cluster galaxies, thereby giving rise to the cutoff at the
bright-end of the galaxy luminosity function and also regulating the
cooling flows.

Gas in the ICM cools via X-ray emission. In the centers of some
clusters, the high density leads to significant loss of energy, such
that the gas radiates away all its energy in a short ($\ll 1/H_0$)
time. In the absence of any heating mechanisms, in order to support
the overlying gas and restore hydrostatic equilibrium, there is a
steady inflow of gas towards the cluster center, which is often
referred to as the classical {\it cooling flow} model
\citep{Fabian1994}. These so-called cool-core clusters~(CC) have
centrally peaked X-ray surface-brightness profiles implying gas
cooling times orders of magnitude shorter than the age of the
cluster. However, (1)~the high resolution $\XMM$ RGS spectra of CC
clusters have not found the expected amounts of cool gas in their
cores
\citep[e.g.][]{Tamura2001,Peterson2001,Kaastra2001,Peterson2003,Xu2002,Sakelliou2002,Sanders2008},
and (2) even though the cooling of the ICM is manifested in the form
of on-going star formation observed in the brightest cluster galaxy of
several clusters \citep[e.g.][]{Mittaz2001,Allen1995}, it is far below
the predicted amount of the star formation rates and CO
\citep[e.g.][]{McNamara1989,Edge2003}. Additionally, the gas
temperature in the central regions as determined from the X-ray
spectra of these clusters is much higher than that expected based on
the cooling flow model and has been found to drop not much below 40\%
the ambient temperature \citep[e.g.][]{Hudson2008}.

Several heating strategies have been proposed to overcome the cooling
flow problem. Feedback from supernovae is an important form of heating
but has been shown to be sufficient to balance energy losses only in
low-luminosity ellipticals with shallower gravitational potentials
\citep{Mathews2003}.

Another heating scenario is thermal conduction which leads to an
inward heat flow from the outskirts of the galaxy clusters.
\citet{Voigt2004} have shown that even though thermal conduction may
provide enough heating to offset cooling in the
hotter~(T$\gtrsim$5~keV) part of the clusters, the central parts of
the cooling region remain largely unaffected by this process. Similar
to supernovae heating, thermal conduction also has the effect of only
slowing down the evolution of intracluster medium by causing the
cooling time to increase by a factor of a few \citep{Pope2005} but
leaves the cooling catastrophe inevitable.

In this work, we focus on the self-regulated AGN feedback as the
current favored mechanism to explain the dearth of cooling by-products
in galaxy clusters
\citep[e.g.][]{Voit2005,Roychowdhury2004,Churazov2002,Binney1995}. In
this framework, accretion of the cool collapsed intracluster
medium~(ICM) ignites the central active galactic nucleus, which
returns a fraction of the accreted power back to the ICM. The bulk of
the energy transfer is believed to happen through mechanical
dissipation of the AGN power. The lead evidence comes from the
observations of numerous galaxy clusters featuring X-ray deficit low
density regions, known as cavities. Such cavities have been observed
to correlate spectacularly with radio jets and lobes indicating that
they are likely regions emptied of ICM by the expanding radio lobes
[e.g.~Perseus, \citet{Boehringer1993}; Hydra-A, \citet{McNamara2000};
A2052, \citet{Blanton2001}; A2597, \citet{McNamara2001}; A4059,
\citet{Heinz2002}; A478, \citet{Sun2003}; A2029, \citet{Clarke2004};
A2199 \citet{Gentile2007}]. The AGN-blown cavities transfer heat to
the ICM potentially by generating sound and weak shock waves
\citep{Jones2002,Fabian2003,Mathews2006}, by doing $p \st d V$ work
against the ambient medium and dissipation of cavity enthalpy in the
wake of buoyantly rising cavities \citep[e.g.][]{Ruszkowski2004,
  Birzan2004}. In addition to {\it direct} AGN mechanical heating via
radio bubbles, there are also alternative proposed mechanisms such as
AGN cosmic-ray heating combined with convection
\citep[e.g.][]{Chandran2007} or conduction
\citep[e.g.][]{Guo2008,Voit2008}.

Numerous results over the last couple of decades have confirmed that
radio-loud AGN dwell preferentially in brightest group and cluster
galaxies~(BCGs), as opposed to other galaxies of the same stellar mass
\citep{Anja2007,Best2007,Bagchi1994,Valentijn1983}. It has also been
found in these and other studies that the CC clusters are particularly
conducive for cD galaxies which are radio-loud, even though the
fraction of radio-loud cD galaxies in CC clusters varies from study to
study ranging from 70\% to 95\%
\citep{Burns1990,Edwards2007,Dunn2006}. The spread in the fraction
between different studies can be attributed to the varying selection
criteria used for constructing cluster samples and the use of not so
up-to-date X-ray and radio observations. The latter effect may result
in the same cluster being identified as a CC cluster in some works and
a non-cool-core~(NCC) cluster in others. A few examples being A1650,
which based on $\einstein$ observations has been quoted as a NCC
cluster by \citet{Burns1990}, but which our data, based on
high-resolution $\chandra$ observations \citep{Hudson2008}, clearly
reveal it to have a cool core with a central cooling time of about a
gigayear and a predicted mass deposition rate of about
100~$\ms$~yr~$^{-1}$ \citep[also see][]{Donahue2005}. Similarly, A3158
and A3195 have been identified based on low-sensitivity and
low-resolution $\asca$ data as CC clusters by \citet{Edwards2007}. Our
results imply otherwise; both are merging systems each with a central
cooling time longer than 12~Gyr and the expected mass deposition rates
being consistent with zero. Despite these inconsistencies, most
studies are by and large in agreement with one another and set the
average abundance of radio-loud CC clusters in the local Universe to
around 80\%.

Recent analyses of galaxy clusters have shown that of those CC
clusters which require heating, at least 40\% harbor cavities that
contain sufficient energy to balance the radiative losses
\citep{Rafferty2008,Nulsen2006,Rafferty2006,Dunn2006}. However, the
details of the various heating mechanisms set into motion by the
central AGN are not clear and are issues currently under
investigation. Also, the local conditions in the ICM that lead to a
quasi-steady state of gas deposition onto the central regions, and
presumably onto the supermassive black hole, and the concomitant AGN
heating of the ICM either periodically or continuously remain largely
unknown. A parallel model that is emerging to explain the deviation of
the observed cluster properties, especially the entropy profiles, from
the predictions of the pure cooling model is linked to preheating or
entropy injection at incipient stages of cluster formation, even prior
to cluster collapse \citep[][ and references therein]{McCarthy2008}.
Preheating is entailed by reduction of central densities, hence,
central luminosities, which leads to flat cluster entropy profiles.
This modification along with post cluster-formation processes, namely
radiative cooling and gravitational heating, provides a better match
to the observed entropy profiles of galaxy clusters. Yet, while
preheating may alone account for the differences in entropy profiles
in NCC clusters, catastrophic cooling at small radii in CC clusters
can still not be bypassed. In order to maintain their entropy profiles
at observed levels, one or more additional sources of on-going heating
are required.

In this work, we aim to gain a more comprehensive understanding of the
AGN-regulated cooling and heating. We scrutinize the ways in which AGN
heating is connected to the cooling of the ICM based solely on the
total radio~(synchrotron) output of the AGN. To achieve this goal, we
use a sample of galaxy clusters for which there exist complete radio
and X-ray data. We take our analysis further by examining the scaling
relations between the BCG near-infrared luminosity and cluster
parameters (mass and luminosity). We derive the mass of the
supermassive black hole using the near-infrared bulge luminosity-black
hole mass relation and inspect whether there is a relation between the
black hole mass and AGN radio luminosity. These correlations are made
taking into account the possibility of obtaining different relations
depending on the cool or non-cool type cluster environment. The
improvement over previous analyses lies in the quality of the cluster
sample and of the available X-ray and radio observations.

We describe the sample in Section~\ref{sample}, giving details about
radio and X-ray data and related quantities in Section~\ref{Radiodata}
and Section~\ref{X-ray data}, respectively. The results are presented
in Section~\ref{results}, including fractions of CC and NCC clusters
with and without central radio sources in Section~\ref{fractions},
cooling and AGN activity in Section~\ref{Cooling AGN activity} and
correlations of the BCG luminosity with radio and X-ray parameters in
Section~\ref{bcg}. We discuss our results in Section~\ref{discussion}
and end with conclusions in Section~\ref{conclusions}. Throughout this
paper, we assume the $\Lambda$CDM concordance Universe with $H_0 =
h_{71}$71~km~s$^{-1}$~Mpc$^{-1}$, $\Omega_{\st m} = 0.27$ and
$\Omega_{\Lambda} = 0.73$.

\section{Our sample}
\label{sample}

The goal of this study is to cross-correlate the cooling activity with
the presence of a radio galaxy\footnote{We use the terms "radio
  galaxy" and "AGN" interchangeably throughout this paper.}  cospatial
with the peak of the cooling flow region or, synonymously, the peak of
the X-ray emission. Further, we want to investigate whether there is a
special coupling between the AGN activity and its surrounding hot
cluster medium in CC clusters as compared to NCC clusters. This calls
for an objectively selected sample. Samples selected based on what is
available in public archives are subject to unknown selection effects
(``archive bias''). For example, the fraction of CC clusters in the
$\chandra$ archive may be biased higher (or lower) than the same
fraction of clusters in the $\XMM$ archive. To this end, we conduct
our study based on the largest X-ray flux-limited sample, the
$\hiflux$ \citep{Reiprich2002} sample, selected from the \rosat
All-Sky Survey outside the Galactic plane using the flux limit,
$f_{\textrm x}~(0.1-2.4)$~keV$ \ge 2 \times
10^{-11}$~erg~s$^{-1}$~cm$^{-2}$.  This sample comprises the 64 X-ray
brightest clusters and spans a redshift range $0.0037 \leq z \leq
0.2153$ with the mean $\left<z\right>\sim 0.05$. All 64 clusters have
been observed with $\chandra$ and all but one~(A2244) have been
observed with $\XMM$ to acquire high quality X-ray data. In this paper
we make use only of the $\chandra$ analysis because we are interested
in the cluster cores and $\chandra$ has currently the best spatial
resolution ($0.^{''}5$) of any X-ray telescope. We note that since
flux-limited samples are as such biased towards clusters with high or
boosted luminosities, this sample might seem to favor CC clusters
preferentially over NCC clusters. At any given redshift, CC clusters
are more likely to be picked up than the NCC clusters due to their
enhanced central luminosities. However, merging clusters present the
same bias as cluster merger events entail temporal enhancements in the
global luminosities and temperatures \citep{Ricker2001}. Since merging
clusters are mostly NCC clusters \citep{Hara2006,Hudson2008}, this may
balance out the former selection bias towards, at least, the {\it
  strong} (to be defined in Section~\ref{fractions} ) CC clusters. We
note that even in the presence of a bias against transition clusters
(neither strong CC clusters nor NCC), objectively selected samples,
such as $\hiflux$, can be directly compared to simulated flux-limited
samples, at both low-$z$ and high-$z$, and the bias may be calculated.

\subsection{Radio data}
\label{Radiodata}

We compiled and in many cases reanalyzed radio observations of all 64
clusters from either literature or archives ($\vla$, NVSS, VLSS and
MOST) to study the radio properties of the centrally located AGN in
the $\hiflux$ clusters. The data from the archives were processed in
the Astronomical Image Processing Software~(AIPS) package provided by
NRAO.

High-frequency archival radio data (500~MHz) were analyzed using the
standard data-reduction procedures within AIPS, wherein the resulting
map usually constituted of a single hybrid image.  Low-frequency data
(330~MHz and 74~MHz), in addition, were carefully analyzed to remove
bad data affected by radio frequency interference~(RFI) using the AIPS
tasks, SPFLG and TVFLG.  In case of pseudo-continuum mode
observations, the effects of bandwidth smearing were tackled by
keeping the data separate over the spectral channels. And lastly, in
order to correct for 3D effects and image degradation due to bright
sources far away from the phase-center, we employed the 3D-imaging
feature embedded in the AIPS task, IMAGR.  This technique of 3D
imaging results in a mosaic of ``facets'', each of which is
independently and simultaneously ``cleaned'' to yield a final map of
the entire primary beam.

\subsubsection{Location of a cluster radio source}
\label{CRS}

\begin{figure}
  \centering
  \includegraphics[angle=-90,width=0.5\textwidth]{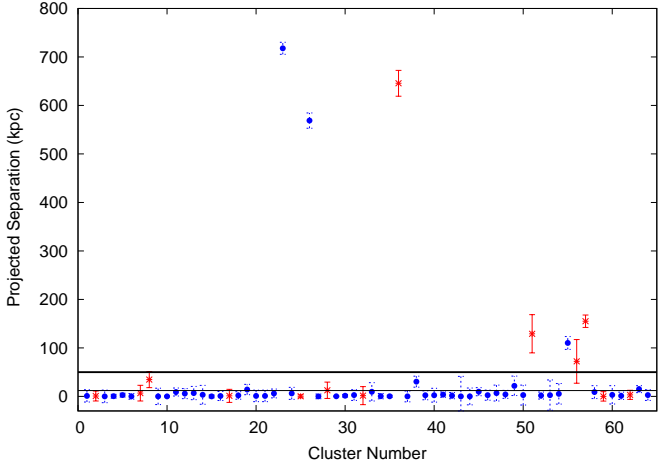}   
  \caption{\small The separation between the BCG and the X-ray Peak.
    Plotted on the X-axis are numbers assigned to each cluster
    arbitrarily. The filled circles (blue) represent BCGs which harbor
    a radio source and the crosses (red) correspond to BCGs without
    one. The black horizontal line at $y = 50 \hinv{-1}$kpc
    corresponds to the criterion for determining whether or not a
    cluster has a CRS and the grey horizontal line at $y =
    12\hinv{-1}$kpc corresponds to the worst uncertainty associated
    with the position of the X-ray peak in the sample.}
  \label{fig:BCG_XP}  
\end{figure}

Before pursuing the cross-correlation of the global X-ray properties
of our sample with the radio properties of cluster central radio
sources~(CCRSs), we need to establish a criterion based on which to
identify central radio sources. Several works have presented evidence
for a special relationship between the cooling activity in cluster
cores and BCGs located within a certain distance to the X-ray peak.
Recently, \citet{Edwards2007} found in their study based on two
samples, the NFPS data set, an X-ray selected sample and the C4
catalog, an optically selected sample built from the SDSS, that only
those BCGs that lie within 70$\hinv{-1}$kpc of the X-ray peak of a
cooling flow cluster have significant line emission. Even though the
optical line emission observed in these BCGs can be inferred either as
a signature of AGN activity or star formation, \citet{Edwards2007}
also show that 74\% of the strongly emitting BCGs in the SDSS sample,
defined as having the $H_{\alpha}$ equivalent width $>2\AA$, have the
diagnostic emitting-line ratios characteristic of AGN activity (with a
likely higher fraction for the BCG). Therefore, the formal basis of
marking a radio source as ``central'' in the study presented here was
to have the AGN within 50$\hinv{-1}$kpc of the X-ray peak, a more
conservative limit than that proposed by \citet{Edwards2007}.  It is
noteworthy that this cut (as opposed to a more stringent cut of
12$\hinv{-1}$kpc, see below) had actually to be invoked only for four
clusters. These four clusters are A3562, A2142, A4038 and A3376 with
the X-ray peak and BCG separation as 30.4$\hinv{-1}$kpc,
21.8$\hinv{-1}$kpc, 14.9$\hinv{-1}$kpc and 14.2$\hinv{-1}$kpc,
respectively.  For the rest of the sample, the flagging was straight
forward in that the separation between the X-ray peak and the radio
active BCG was less than 12$\hinv{-1}$kpc. The 12$\hinv{-1}$kpc as the
yardstick comes from the fact that since the $\hiflux$ clusters span
two orders of magnitude in redshift, the $\chandra$ resolution implies
varying accuracies with which the X-ray peak may be determined for
different clusters and 12$\hinv{-1}$kpc corresponds to the worst
1-$\sigma$ uncertainty.  The separation between the BCG and the X-ray
peak for all 64 clusters is shown is Figure~\ref{fig:BCG_XP}. Applying
this criterion, we find a total of 48 clusters with centrally located
radio sources.

\subsubsection{Integrated radio luminosity of a CCRS}
\label{IRL} 

\begin{figure*}
  \begin{minipage}{0.49\textwidth}
    \centering
    \includegraphics[angle=-90,width=\textwidth]{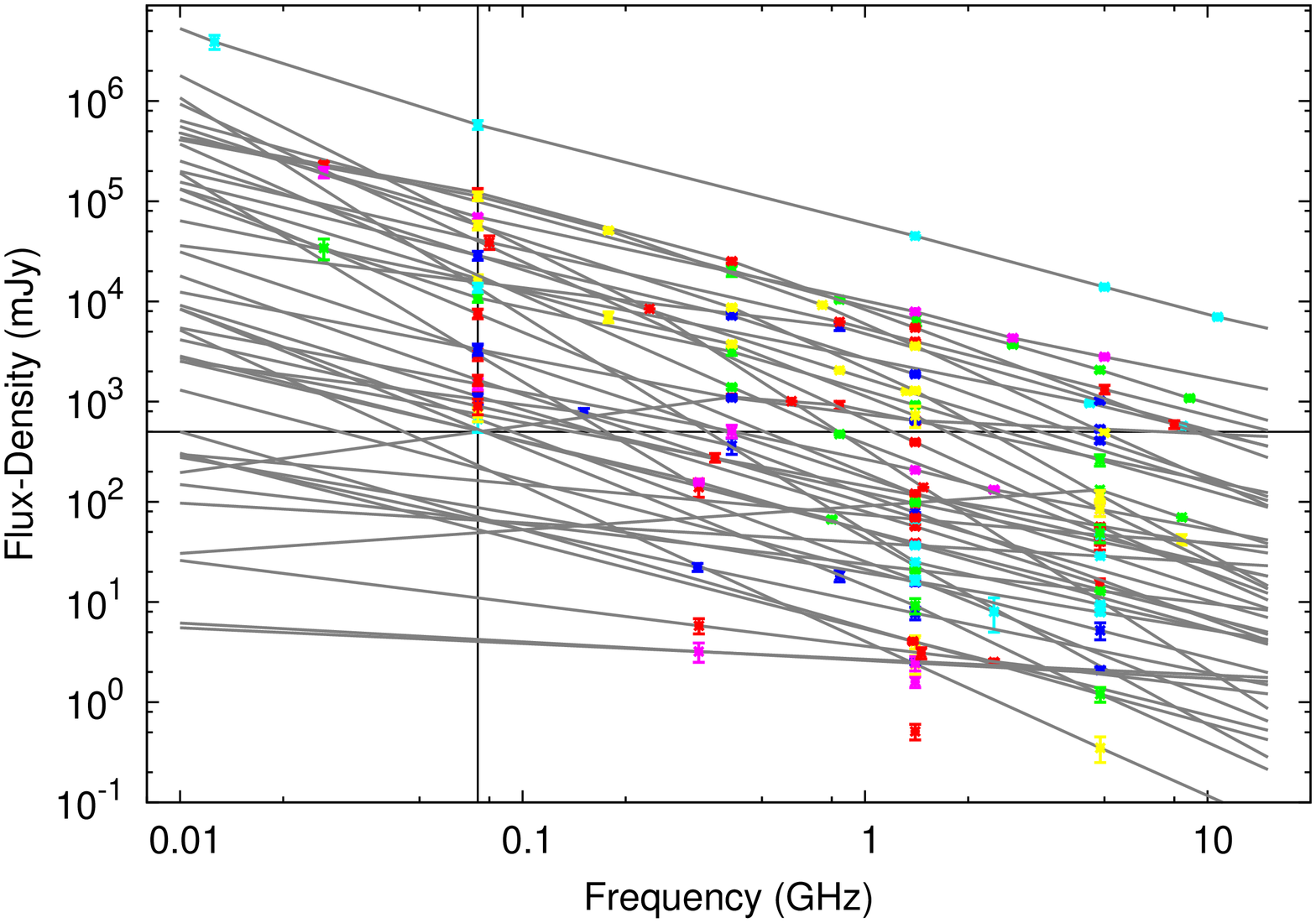}    
  \end{minipage}\hfill
  \begin{minipage}{0.49\textwidth}
    \centering
    \includegraphics[angle=-90,width=\textwidth]{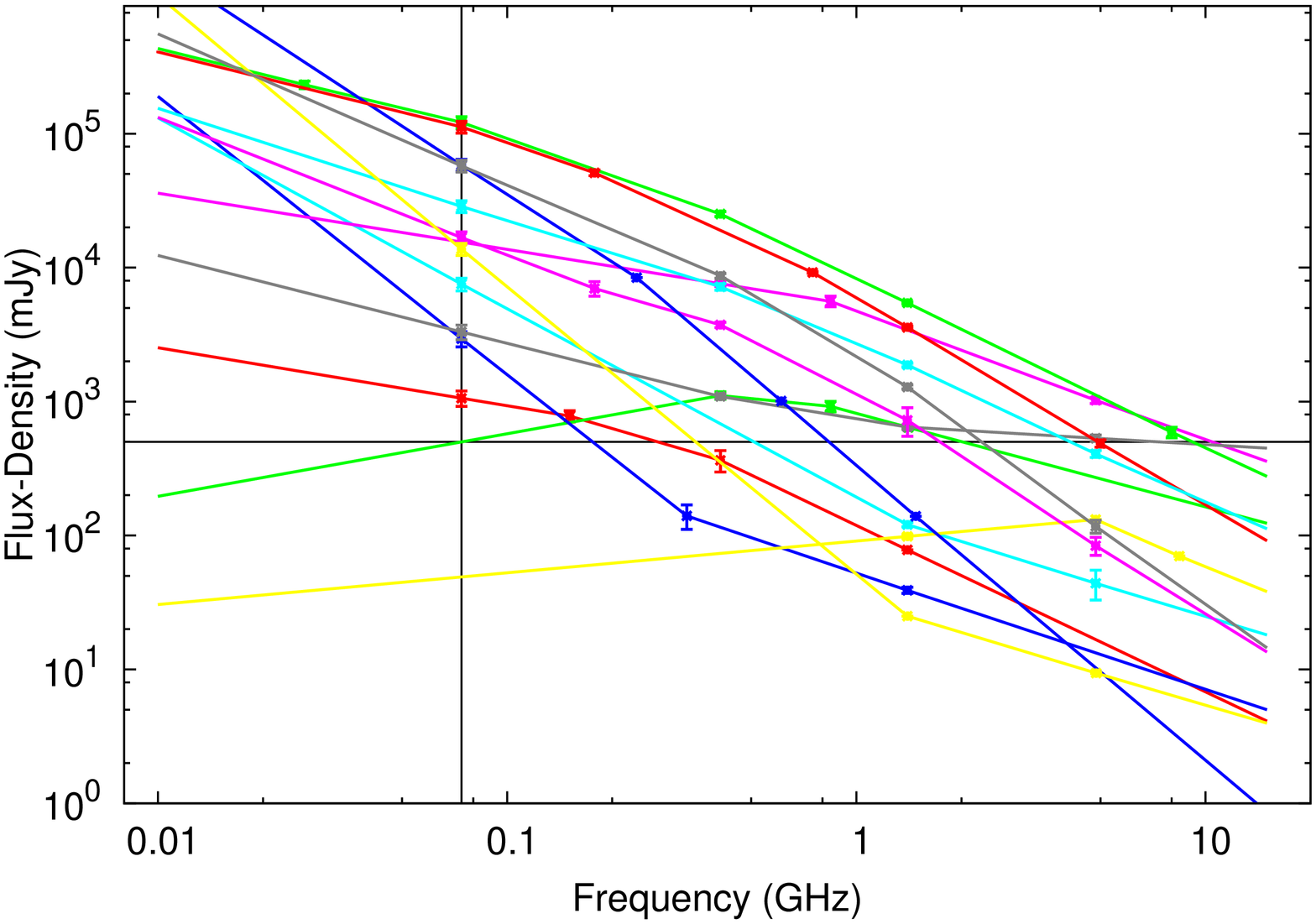}
  \end{minipage}
  \caption{\small The radio spectra of the cluster central radio
    sources. Shown in the left panel are the spectra of all 48 CCRSs,
    where the symbols are the actual measurements. Most of the
    measurements at 74~MHz and 1.4~GHz are taken from VLSS and NVSS,
    respectively. Shown in the right panel are the spectra of only
    those CCRSs which show spectral curvature. The black horizontal
    solid line represents the VLSS point-source sensitivity constraint
    at the VLSS observing frequency of 74~MHz indicated by the black
    vertical line.}
  \label{fig:radio-spectra}  
\end{figure*}

One of the primary concerns in accumulating the radio data was to have
a good spectral coverage, particularly, at the low-end of the radio
spectrum. Of the 48 CCRSs, 65\% have radio data below 500~MHz and 46\%
have radio data below 80~MHz. Low-frequency radio observations are
important to perform a full spectral analysis for these clusters for
two reasons. First, since the main contribution to the synchrotron
radio luminosity comes from the low-end of the radio frequency
spectrum, this will result in precise determination of the synchrotron
power in these systems. In our subsequent study, the energy in the
radio emitting particles will be compared to the mechanical energy of
the cavities, and thereby the partitioning of energy between radiation
and mechanical~(cavity) energy can be measured. Second, with dense
enough spectral sampling, spectral breaks may also be visible (as have
already been seen in a few CCRSs in the $\hiflux$ sample, see
below). A spectral break in a system with cavities is an extremely
useful observable since it is representative of the time since the
last injection event or particle production, and therefore, a good
indicator of the age of the cavity emission. The CCRSs, which we
already know to show spectral breaks are presently being cross-checked
with presence of cavities for a future study.

For sources with no observations or confirmed detection below 100~MHz,
the low-end of their spectra were constrained using the 74~MHz $\vla$
Low-frequency Sky Survey~(VLSS) with an average point-source detection
limit of 0.5~Jy/beam and a resolution of $80^{''}$. Shown in the left
panel of Figure~\ref{fig:radio-spectra} are the spectra of all CCRSs.
Shown in the right panel of Figure~\ref{fig:radio-spectra} are the
spectra of only a subset of CCRSs that show interesting features such
as spectral breaks and turn-overs indicative of spectral aging and
synchrotron self-absorption, respectively. Amongst the CCRSs shown in
the right panel, we note there are also a few clusters which show
spectral steepening at low-frequencies, which we believe is due to the
superposition of different radio components with different spectral
properties (due to varying sizes and distances from the central
engine).

The details of the radio data used for this work, such as the various
frequencies and the corresponding flux-densities used to estimate the
radio luminosities of the cluster central radio sources, along with
the references can be found in Table~\ref{tab:radio_data_ref}. The
synchrotron radiation is assumed to have a powerlaw spectrum given by
$S(\nu) \propto \nu^{-\alpha}$, where $S(\nu)$ is the flux density at
frequency $\nu$. Thus, the integrated rest-frame radio luminosities,
$\lr$, of the CCRSs were calculated by step-wise integration:
\begin{eqnarray}
  L_{i+1} & = & 4 \pi D^2_l \, S_0 \int_{\nu_i}^{\nu_{i+1}}
  (\nu/\nu_0)^{-\alpha_{i+1,i}} d\nu \nonumber \\[8pt]
  & = & 4 \pi D^2_l \, \frac{S_0
    \nu_0^{\alpha_{i+1,i}}}{(1-\alpha_{i+1,i})}  
  \left(\nu_{i+1}^{(1-\alpha_{i+1,i})} -
    \nu_i^{(1-\alpha_{i+1,i})}\right), 
  \label{RadLum}
\end{eqnarray}   
where $S_0$ is the flux density of the radio source at either of the
two rest-frame frequencies, $\nu_i$ or $\nu_{i+1}$, $L_{i+1}$ is the
radio luminosity in the frequency range \mbox{[$\nu_i$, $\nu_{i+1}$]},
$\alpha_{i+1,i}$ is the spectral index between $\nu_i$ and
$\nu_{i+1}$, and $D_l$ is the luminosity distance. The total radio
luminosity was calculated by extrapolating the spectral indices
obtained at the lowest observed frequency to 10~MHz and at the highest
observed frequency to 15~GHz.  Thus, $L_{\st{tot}} = \sum L_{i+1}$. Of
all the clusters with CCRSs, 27 have (reliable) data at more than two
frequencies and 18 have data at two frequencies. The remaining three
clusters~(A576, A3158 and A3562) have data at only one frequency and
we used $\alpha=1$, the average spectral index of the CCRSs in our
sample, to calculate their total radio luminosity.

Even though the formal errorbars for $\lr$ were derived using the
background root-mean-square in the maps, these do not take into
account the uncertainty arising due to the lack of knowledge of the
shape of the radio spectra down to the lowest frequencies, except for
a handful of radio sources well-studied at all radio-frequency
bands~(such as, Hydra-A~(A0780), Centaurus~(A3526), A1795, A2029,
A2052, A2199, A2597 and A4059). Radio sources often show a spectral
turn-over at low-frequencies. This is attributed to synchrotron
self-absorption which kicks in with increasing optical depth and is
manifested by a rising spectrum with $\alpha = -2.5$. Two examples of
spectral turn-overs can be easily seen in the right panel of
Figure~\ref{fig:radio-spectra}. Considering the possibility of other
CCRSs showing similar turn-overs, we calculated a lower-limit on $\lr$
based on the assumption that the spectra of the CCRSs turn over right
below the lowest observed frequency. The difference between the lower
limit derived in this manner and the integrated radio luminosities
assuming the continual of the spectra beyond the lowest observed
frequency provides a more realistic, albeit conservative, uncertainty
on $\lr$. This is the reason for having highly asymmetric errorbars
for $\lr$, as can also be seen in many of the plots.

\begin{figure}
  \centering
  \includegraphics[angle=-90,width=0.5\textwidth]{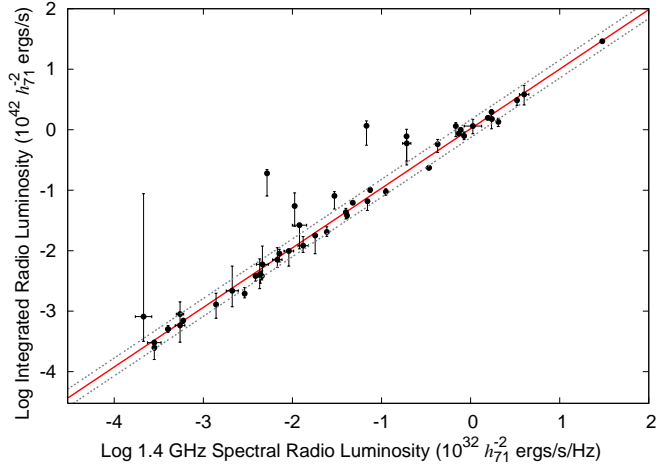}
  \caption{\small A comparison between the radio luminosity integrated
    between 10~MHz and 15~GHz and the monochromatic radio luminosity
    at 1.4~GHz~(red solid line). The dotted grey lines correspond to
    1-$\sigma$ deviation above and below the best-fit line.}
  \label{fig:RadSpecLum}  
\end{figure}

In other works \citep[e.g.][]{Burns1990, Peres1998}, very often the
comparisons between the quantities representing the cooling flow
strength in clusters and the radio power of the BCGs are based solely
on the monochromatic radio luminosities, such as, the 1.4~GHz
luminosities derived from NVSS or FIRST or 5~GHz luminosities derived
from the Green Bank Survey. A useful exercise followed up with our
data was to compare the integrated radio luminosities to
$L_{1.4~\textrm{\small GHz}}$, where the latter is given by
\begin{equation}
L_{1.4~\textrm{\small GHz}} = 4 \pi D^2_l \, S_{1.4~\textrm{\small GHz}}
  (1+z)^{\alpha_{1.4, i} - 1} \; ,  
\label{eqn1}
\end{equation}

where $\alpha_{1.4, i}$ is the spectral index between 1.4~GHz and one of the
two neighboring frequencies. Shown in Figure~\ref{fig:RadSpecLum} is the
integrated radio luminosity versus the monochromatic radio luminosity at
1.4~GHz. Also shown is the best-fit powerlaw derived using the {\it FITEXY}
least-squares line-fitting routine \citep{Press1992} given by
\begin{equation}
  \frac{\lr}{10^{42}~ h_{71}^{-2}~~\st{ergs~s}^{-1}} = \st a
  \times \, \left(\frac{L_{1.4~\st{\small GHz}}}{10^{32}~h_{71}^{-2}
  ~~\st{ergs~s^{-1}~Hz^{-1}}}\right)^{\st b} \,  ,
\label{eqn2}
\end{equation}
where $\st a=1.04 \pm 0.03$ and $\st b=0.98 \pm 0.01$. This algorithm
allows for fitting in only one parameter ($\lr$ in this case) but
takes uncertainty in both $X$ and $Y$ into account. Our study shows
that there is a fairly good correlation between the $\lr$ and
$L_{1.4~\textrm{\small GHz}}$.  This is not surprising since once a
pedestal value for radio sources is determined, the total power should
scale with the spectral index, which for the CCRSs in our sample is
quite similar and centers around unity. There are, however, a few
CCRSs (such as 2A0335, A3376, MKW3S and A4038) which are inconsistent
with the best-fit relation, all of which have spectral-indices steeper
than unity at low frequencies and were excluded while determining the
best-fitting powerlaw. Hence, we conclude that even though the
monochromatic luminosity (in this case at 1.4~GHz) is a good proxy for
the total radio luminosity and may be used in cases where additional
spectral information is not available, for precise radio correlations
demanding least intrinsic scatter, the total radio power computed from
detailed spectral analyses should be used.

\subsection{X-ray data}
\label{X-ray data}

The complete $\hiflux$ sample has $\chandra$ observations and the data
have been homogeneously reprocessed using CIAO~3.2.2 and
CALDB~3.0. For a detailed description of the X-ray data-reduction, and
data- and error-analysis, the reader is referred to
\cite{Hudson2008}. In the following we briefly describe only those
cluster X-ray parameters that are meaningful in the context of the
present work.

\subsubsection{Cluster parameters}
\label{cluster parameters}

We extracted the central density profiles by fitting either a single
or double $\beta$- model to the surface-brightness profile annuli.
Similarly, the central temperature profiles were derived by fitting
spectra to annuli to an absorbed thermal model. From the temperature
and density profiles, the central cooling time at $r = 0.004
\rvir$~(0.4\%~$\rvir$), $\ct$, entropy, $K$, and cuspiness, $\alpha$,
were calculated as per the following expressions:

\begin{eqnarray}
  \ct  = \frac{5}{2} \frac{(n_{\st i0} + n_{\st e0} ) kT_{48}}{n_{\st e}^2
    \Lambda (T_{48})} \quad ; \quad 
  K(r) = kT(r)n_{\st e}^{-2/3}(r) \nonumber \\
  \alpha = - \frac{d \st{log}(n_{\st e})}{d \st{log}(r)} 
  \label{eqn3}  
\end{eqnarray} 
where $n_{\st i0}$ and $n_{\st e0}$ are the central ion and electron
densities, respectively, determined at $r = 0.004 \rvir$, $T_{48}$ is
the average temperature of the $0-0.048\rvir$ region, $\Lambda
(T_{48})$ is the cooling function for a plasma at $T_{48}$ and $n_{\st
  e}(r)$ is the electron density at a given radius $r$ from the
cluster center. The cuspiness is calculated at a distance $r = 0.04
\rvir$ from the cluster center. Here, $\rvir$ is the radius within
which the average cluster mass density is 500 times higher than the
critical density of the Universe.

To derive the cluster entropy profiles, the best fit density profiles
were binned in steps of $2^{''}$. Then for each bin, the value of
temperature corresponding to that radius was adopted to calculate the
entropy for that bin using the expression in
Equation~\ref{eqn3}. Since the annuli created to derive the
temperature profiles need not necessarily coincide with the $2^{''}$
density bins, in the case where there was a jump in the temperature
within a density bin, the average value of the two temperatures was
used.

The virial temperature of the cluster, used as a scaling parameter in some of
the cross-correlations presented in the forthcoming sections, was determined
by fitting the temperature profiles to broken powerlaws.  This was done so as
to prevent the cool-core gas from biasing the estimate for the global `viral'
temperature, $\tvir$. In those cases, where the inner powerlaw had a declining
slope towards the center, which is representative of the cool gas at the
centers of CC clusters, we excluded the core region as determined from the
break in the powerlaw from the fit. From the estimate of $\tvir$, the virial
mass, $\mvir$, within $\rvir$ was determined using the relation by
\cite{Finoguenov2001}, $\mvir = \st{a} \, k \tvir^{\st b} \, 10^{13} \ms $,
where $\st a=2.5 \pm 0.2$~h$_{71}^{-1}$ and $\st b=1.676\pm0.054$.

\begin{figure}
  \centering
  \includegraphics[angle=-90,width=0.5\textwidth]{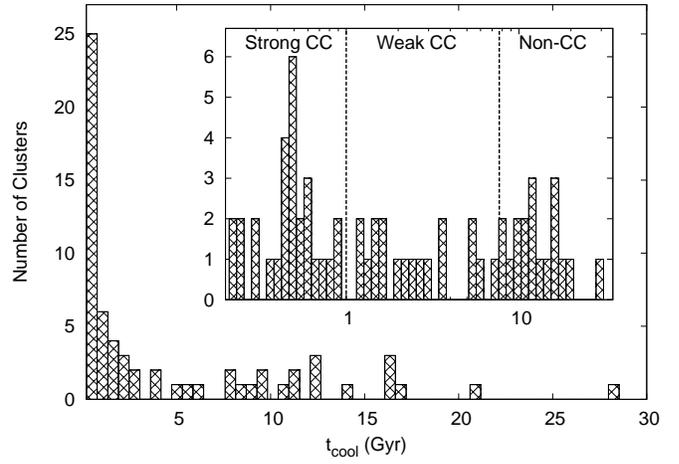}   
  \caption{\small The central cooling-time distribution clearly
    showing the steep rise in the fraction of clusters with cooling
    times shorter than 1~Gyr. For clarity, we show the distribution in
    both linear as well as log scale (inset panel) on the x-axis. }
  \label{fig:CThist}  
\end{figure}

\begin{figure*} 
  \begin{minipage}{0.5\textwidth}
    \centering
    \includegraphics[angle=-90,width=\textwidth]{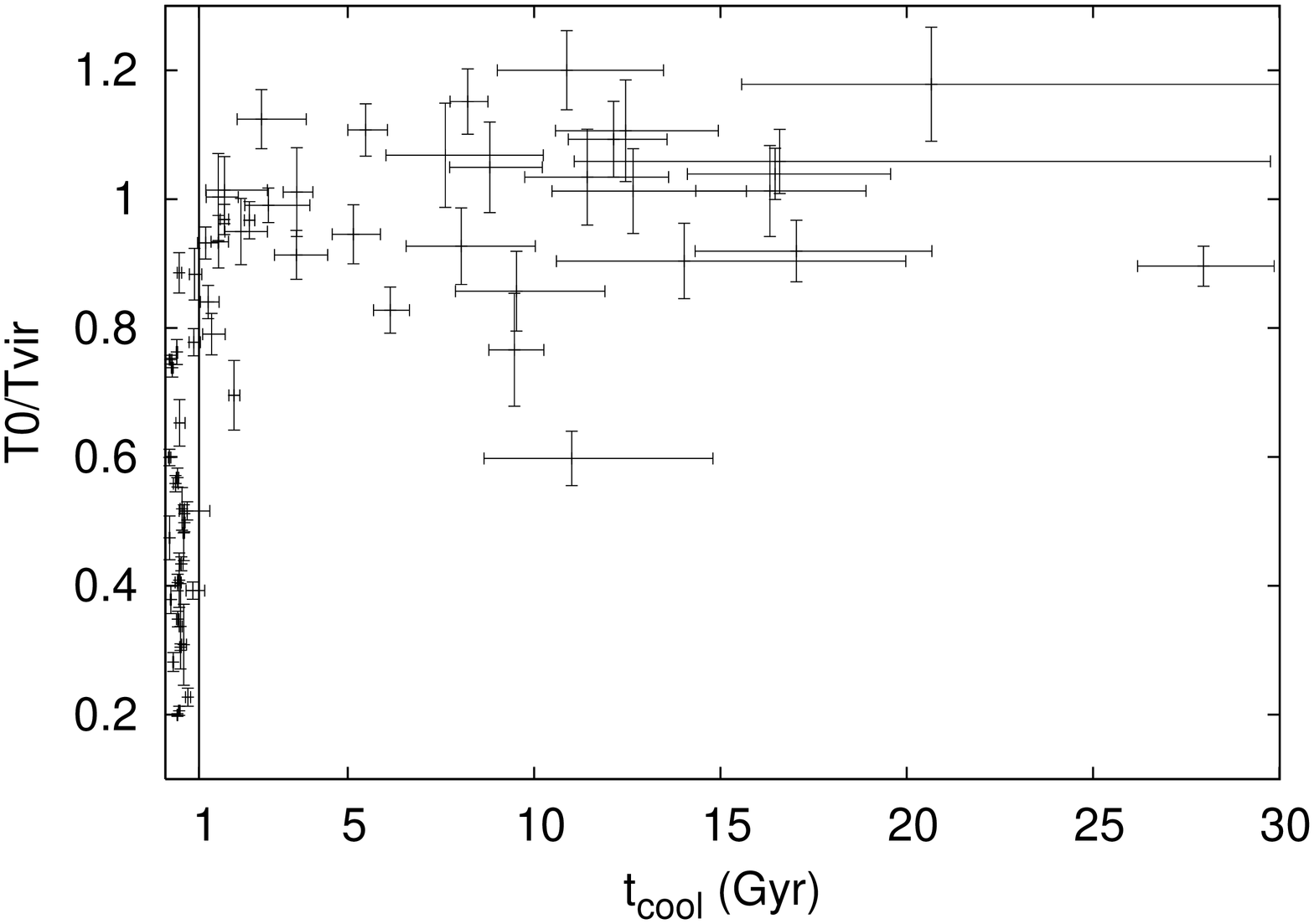}    
  \end{minipage}\hfill
  \begin{minipage}{0.5\textwidth}
    \centering
    \includegraphics[angle=-90,width=\textwidth]{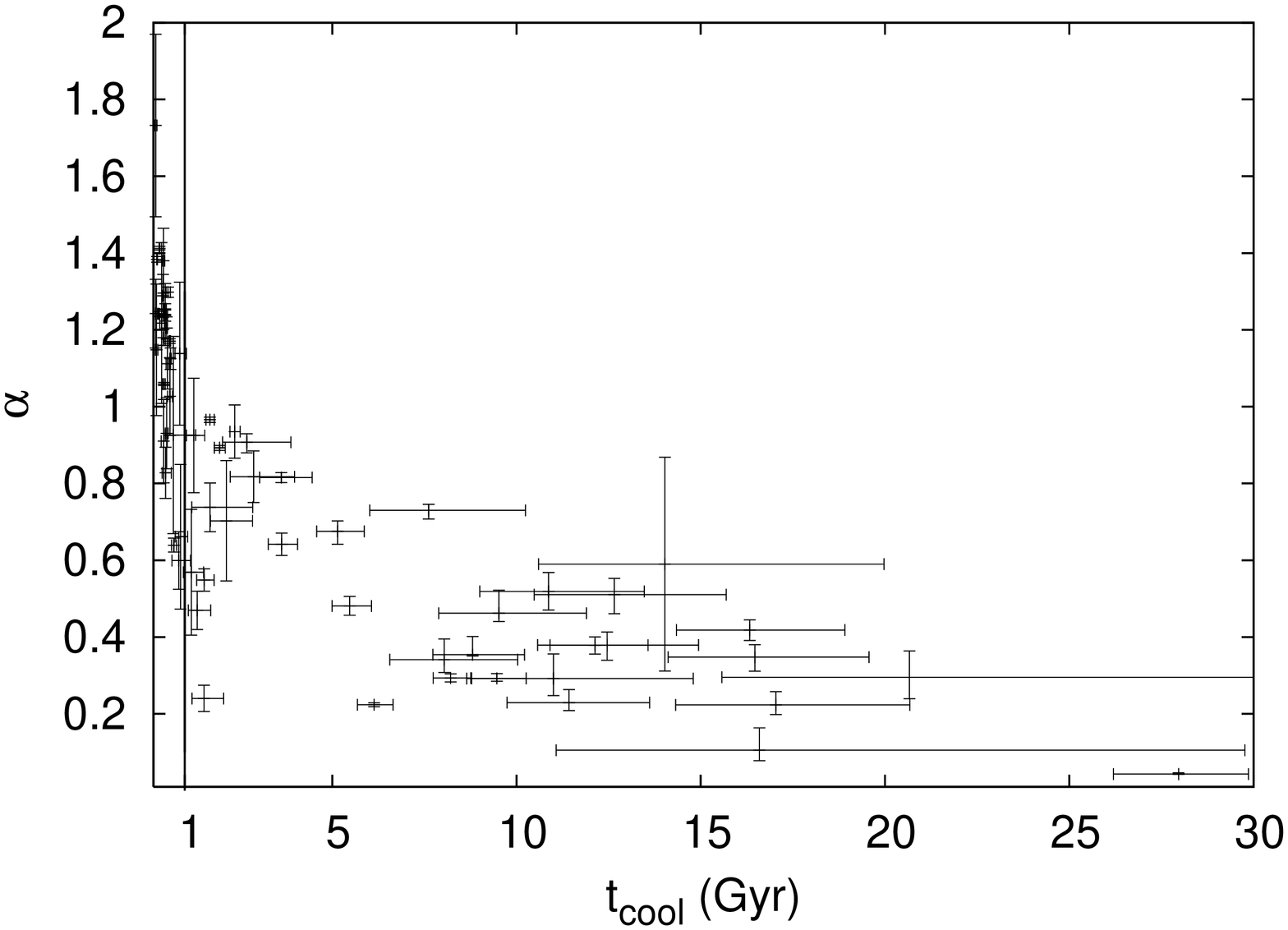}   
  \end{minipage}
  \caption{\small Cooling time as an indicator of cool core. {\it
      Left:} The central temperature drop. {\it Right:} The slope in
    the surface brightness profile, {\it cuspiness}, at $4\%$ of
    $\rvir$.}
  \label{fig:1Gyr}  
\end{figure*}

The classical mass deposition rates, $\mdr$, were derived from the gas
temperature and density profiles. $\mdr$ is the ratio of the total gas
mass within the cooling radius, $\rc$, defined as the region at which
the $\ct < 7.7$~Gyr, to the cooling time at this radius. Hence, $\mdr
(r) = M_{\st{gas}}(r < \rc)/\ct(\rc)$ is a measure of the rate at
which the mass should be dropping out of the X-ray band provided there
is no source of heating. As the main focus for this work is on the
central parts of clusters where cooling is most dominant, the X-ray
analysis was performed only for the core
regions.

\section{Results}
\label{results}

\subsection{CC and NCC cluster fractions: With and without a central
  radio source} 
\label{fractions} 

A well-known problem related to cooling-flows has been that of
choosing an apt diagnostic for determining a cool-core cluster. To
separate out the cool-core clusters from the non-cool-core ones,
\cite{Hudson2008} searched for a bimodality in several of the X-ray
observed and derived quantities, such as, the central cooling time,
$\ct$, the cooling radius defined as the radius out to which $\ct <
7.7$~Gyr, the central entropy, the central density, the central
luminosity, the mass deposition rate, the central temperature drop and
the slope in surface brightness profile, wherein ``central'' refers to
$0.004 \rvir$~($0.4\%$ of $\rvir$). Based on the K-Mean Method~(KMM)
algorithm \citep{Ashman1994} as a test for bimodality (or
tri-modality) in the parameters, \cite{Hudson2008} found $\ct$ as the
best measure for cooling to divide the CC and NCC clusters.

The $\ct$-distribution histogram shown in Figure~\ref{fig:CThist}
shows a peculiarity in that the distribution is marked by a sharp
increase at $\ct < 1$~Gyr; the fraction of clusters with $\ct < 1$~Gyr
being 44\%.  This oddity at $\ct < 1$~Gyr is also visible in two other
properties, (1)~the central temperature drop shown in the left panel
of Figure~\ref{fig:1Gyr}, defined as the ratio of the temperature in
the innermost region to the virial temperature and (2) the slope in
the surface brightness profile, the 'cuspiness', shown in the right
panel of Figure~\ref{fig:1Gyr}. Both the quantities show a break
around a central cooling time of $\ct < 1$~Gyr, even though the
decrease in the central temperature drop is much more pronounced than
the increase in the cuspiness. This result is also in concert with the
recent findings of \citet{Rafferty2008}, who investigated the relation
between star formation, cooling activity in the ICM and AGN heating,
based on a sample of 47 cluster center dominant galaxies~(CDGs).
According to their study, only the CDGs with cooling times below
$0.8$~Gyr exhibit positive color gradients, signifying an increase in
star formation with decreasing distance from the galaxy center. The
underlying reason for this behavior is not yet clear. It may be that
the cool gas at the centers of galaxy clusters is feeding the star
formation, in which case the short cooling times should be tied with
the time-scale over which the stars form and their light declines. It
may well also be that AGN activity at the center of mass flows
triggers star formation at the central regions. If the AGN feedback is
responsible for regulating the cooling flows in clusters, then the
cooling time-scale may possibly reflect an intimate link with the time
scale of the AGN outbursts (Section~\ref{discussion}).

Based on the above results, we divided our sample into three
categories, (1)~strong cool-core~(SCC) clusters with $\ct < 1$~Gyr,
(2)~weak cool-core~(WCC) clusters with $1$~Gyr$< \ct < 7.7$~Gyr (the
upper limit of 7.7~Gyr is the usually assumed value for the cooling
time corresponding to $z=1$, signifying the lookback time since the
last major heating event, see McNamara \& Nulsen 2007)
\nocite{McNamara2007} and (3)~non-cool-core~(NCC) clusters with $\ct >
7.7$~Gyr. These cuts result in 44\% SCC clusters, 28\% WCC clusters
and 28\% NCC clusters. The need to divide the distribution into three
subgroups is bolstered by the fact that the KMM test showed adding a
third sub-group improved the likelihood ratio, giving rise to a
tri-modal distribution.

\begin{figure}
  \centering
  \includegraphics[angle=-90, width=0.35\textwidth]{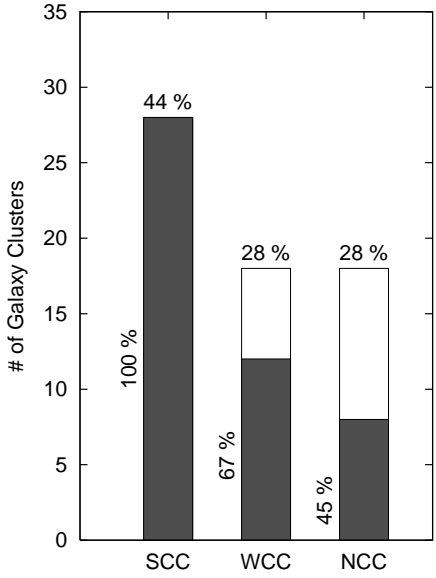}
  \caption{\small The fraction of strong cool-core~(SCC) clusters,
    weak cool-core~(WCC) clusters and non-cool-core~(NCC)
    clusters in the $\hiflux$ sample. Also shown are the fraction of
    clusters containing central radio sources for each
    category~(shaded).}
  \label{fig:Fraction}  
\end{figure}
\begin{figure}
  \centering
  \includegraphics[angle=-90,width=0.5\textwidth]{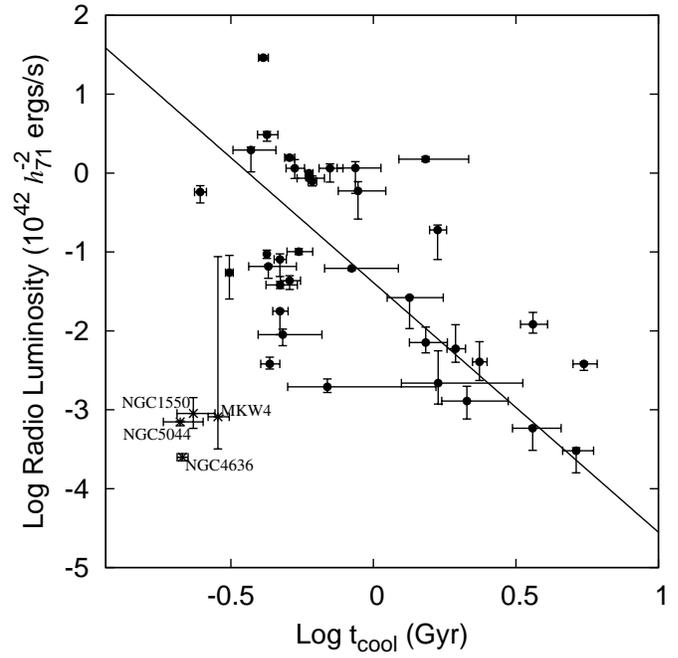}    
  \caption{\small The central cooling-time versus the integrated radio
    luminosity for the CCRSs in CC clusters (SCC $+$ WCC). The black
    solid line represents the anti-correlation trend which breaks down
    for clusters with $\ct <1$~Gyr. The labeled clusters are outliers
    with peculiar properties (see text for more).}
  \label{fig:CorrPlots1}
\end{figure}

On cross-correlating the clusters with the presence of a CCRS, we find
that \textit{all} SCC clusters show cluster-center radio sources
(Figure~\ref{fig:Fraction}). The resulting fraction of CCRSs amongst
WCC clusters is 67\% and that in NCC clusters is 45\%. A
non-negligible fraction of CCRSs in the WCC and NCC cluster population
makes it uncertain whether there is a fundamental one-to-one
correspondence between AGN heating and the lack of the expected
cluster cooling. On the other hand, the probability of a BCG
manifesting AGN activity clearly increases with decreasing cooling
time. The next question that then arises is whether the radio
luminosity of the central cluster radio source itself is correlated
with $\ct$. This is shown in Figure~\ref{fig:CorrPlots1} for SCC
clusters and WCC clusters. From hereon, we refer to the combined set
of SCC and WCC clusters as the cool-core~(CC) clusters. This plot does
not present a straight-forward interpretation of the interdependence
between the AGN synchrotron power and the cooling time-scale. As a
whole, there seems to be an anti-correlation between the two
quantities but this seems to break down for clusters with $\ct
\lesssim 1 $~Gyr. This apparent anti-correlation could be indicative
of a need for more powerful AGN as heating agents in clusters with
short $\ct$. Yet the absence of any correlation between the $\ct$ and
$\lr$ at short cooling times ($<$~Gyr) implies that the AGN luminosity
is more sensitive to a physical quantity other than the gas cooling
time, such as possibly the mass deposition rate, $\mdr$ (see
Section~\ref{Cooling AGN activity}).

Shown as crosses in Figure~\ref{fig:CorrPlots1} are four systems,
NGC4646, NGC1550, NGC5044 and MKW4, which clearly depart from this
trend and all of which are groups. A general property that the groups
in our sample seem to share is that apart from having low temperatures
(both virial and central), they also all have high central densities
and subsequently short $\ct$ (see Eq.~\ref{eqn3}). On the other hand,
the groups tend to have very steep density gradients resulting in
small classical mass deposition rates, $\mdr$. In other words, $\mdr$
is more sensitive to the mass encompassed within the integration
radius (see Section~\ref{cluster parameters}) than the cooling time at
that radius [also see Figure~6(G) of Hudson~et~al.~2008].  The
behavior of $\lr$ versus $\mdr$ is investigated in
Section~\ref{Cooling AGN activity}. The fourth outlier, MKW4, is an
interesting cluster under intensive study at radio wavelengths (see
Section~\ref{bcg2}). Assuming the anti-correlation interpretation is
correct, the best fit powerlaw excluding the four outliers derived
using the bisector linear regression routine, {\it BCES} from
\cite{Akritas1996a} is
\begin{equation}
  \frac{\lr}{10^{42}~ h_{71}^{-2}~~\st{ergs~s^{-1}}} = (0.041 \pm 0.016)
  \times \, \left(\frac{\ct}{\st{Gyr}}\right)^{-3.16 \pm 0.38}  \,  .
\label{eqn4}
\end{equation}
This routine, like {\it FITEXY}, includes uncertainties in both the
quantities but also additionally performs the minimization in both the
dimensions. The Spearman rank correlation coefficient is $-0.63$ and
the probability for the null-hypothesis is $8 \times 10^{-6}$.

For comparison with other works, we also determined the fraction of
CCRSs in CC clusters, the fraction being 87\%. This is consistent with
the result of \citet{Dunn2006}, who analyzed a low-redshift sample of
clusters (B55) selected from pre-$\rosat$ data. Even though they find
a slightly higher fraction (95\%) of CC clusters with CCRSs, they used
a lower cut in $\ct$ to determine CC clusters and, additionally,
selected only those clusters which showed a central temperature drop
$>2$. Using these criteria reduces the fraction of CC clusters in our
sample to $25\%$ but increases the fraction of CCRSs in CC clusters to
$100\%$. Similarly, \citet{Burns1990} finds a somewhat lower fraction
of $70\%$ but the classification into CC and NCC clusters therein is
based on the Hubble time. Using the Hubble time as the cut in $\ct$
increases the fraction of CC clusters in our sample to $89\%$ and
reduces the fraction of CCRSs in CC clusters to $78\%$. We also bear
in mind that the result by \citet{Burns1990} is based on an incomplete
sample and old X-ray $\einstein$ data. Furthermore, the radio data
used by \citet{Burns1990} are based on monochromatic 5~GHz VLA
observations sensitive to largest structures of only about an
arcminute, which in some cases might lead to over-resolved structures
and, hence, an under-estimation of the radio luminosity.

\subsection{Cooling and AGN activity}
\label{Cooling AGN activity}

\begin{figure*}
  \begin{minipage}{0.49\textwidth}
    \centering
    \includegraphics[angle=-90,width=\textwidth]{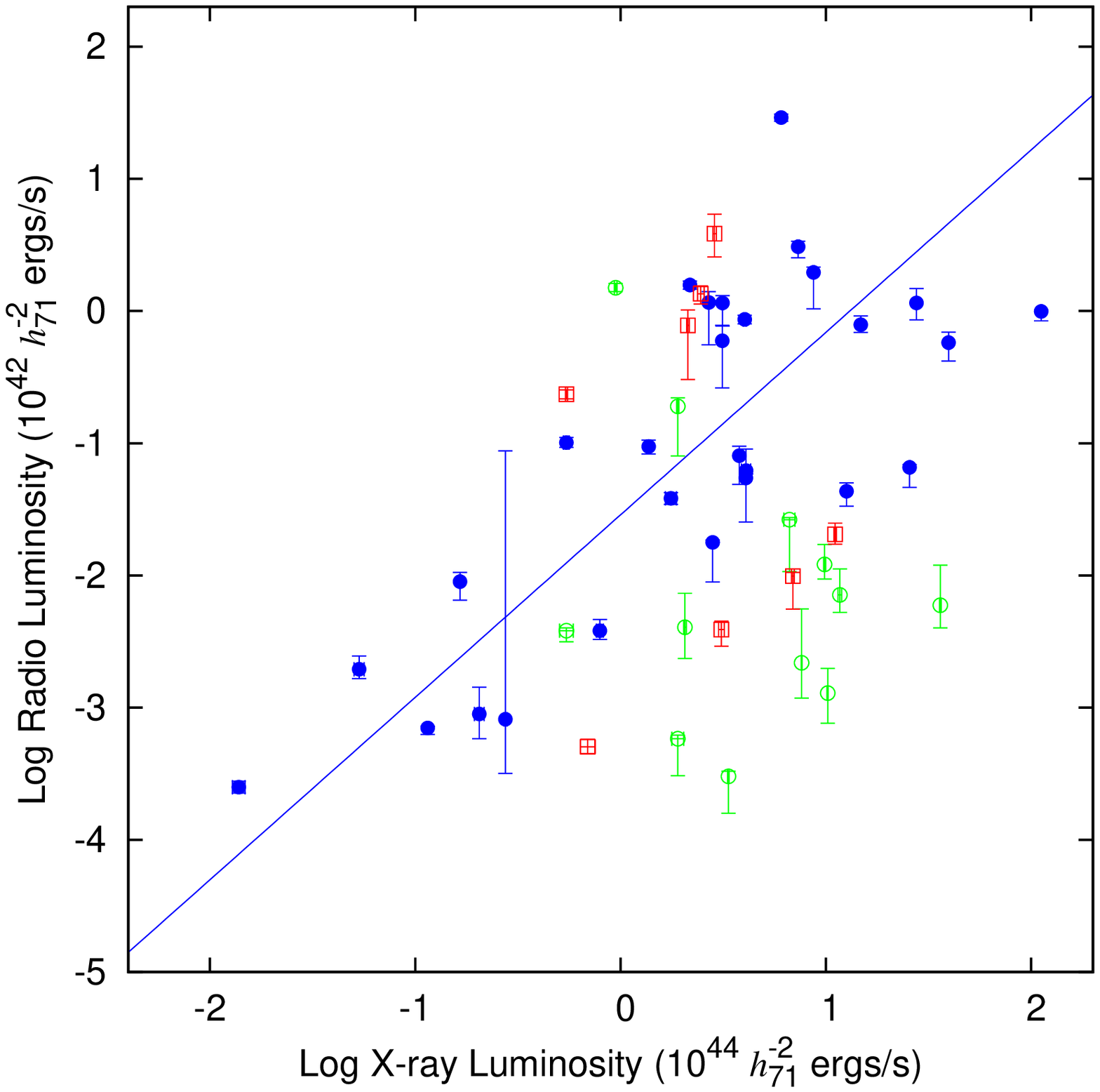}    
  \end{minipage}\hfill
  \begin{minipage}{0.49\textwidth}
    \centering
    \includegraphics[angle=-90,width=\textwidth]{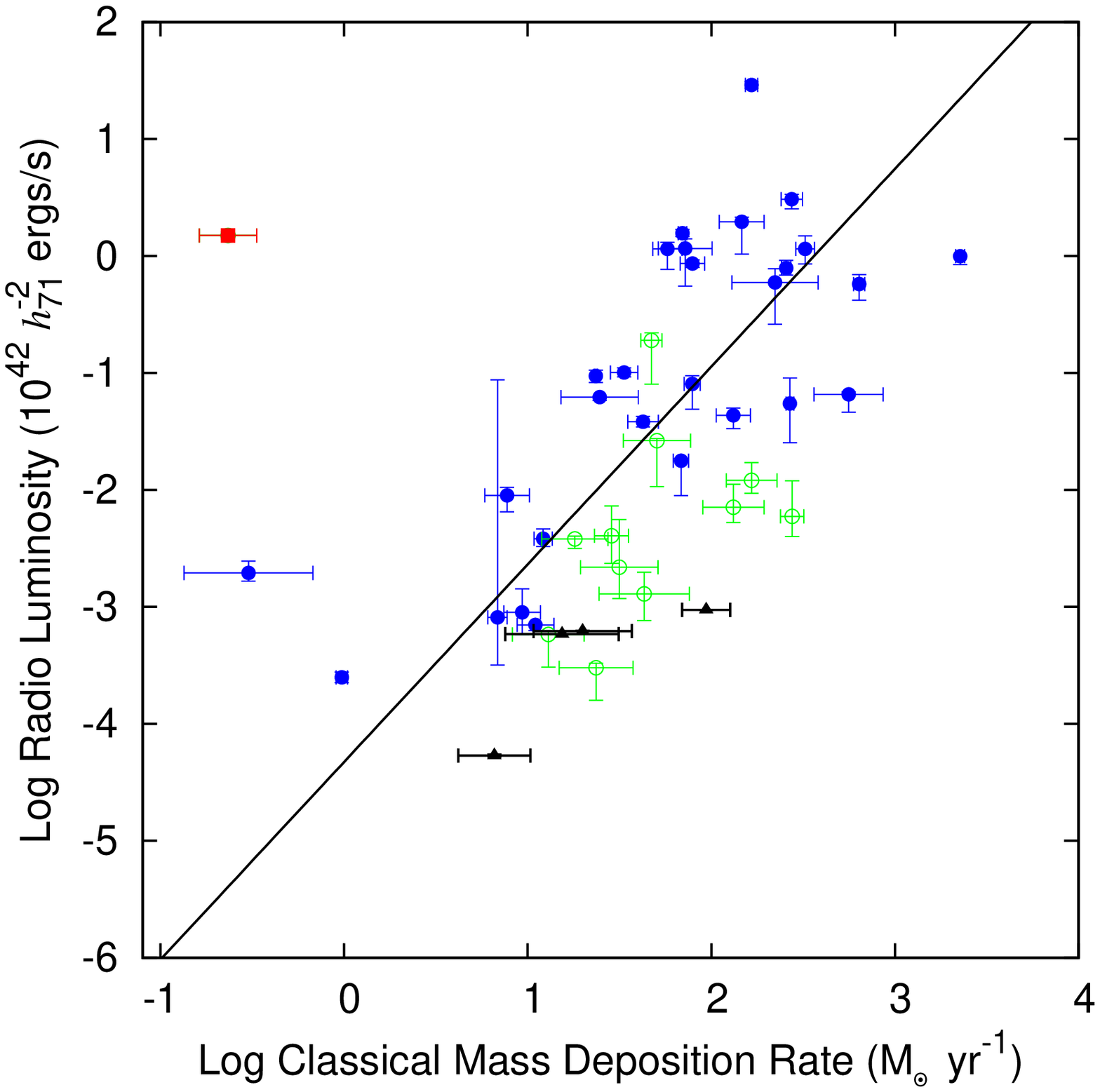}
  \end{minipage}
  \caption{\small Radio and X-ray correlation plots. \textit{Left}:
    Total radio luminosity vs. $\rosat$ bolometric X-ray luminosity
    for SCC (filled blue circles), WCC (open green circles) and NCC
    (open red squares) clusters.  \textit{Right}: Total radio
    luminosity vs. the classical mass deposition rate for SCC clusters
    (filled blue circles) and WCC clusters (open green circles). The
    black triangles are four WCC clusters with $\mdr > 1$ and no
    central radio source. Hence, these are only upper limits. The
    filled red square on top left is A2634, also a WCC cluster. The
    solid black line is the best fit through CC clusters (SCC and WCC
    clusters combined) excluding A2634.}
  \label{fig:CorrPlots2}
\end{figure*}

We looked for correlations between the radio luminosity of the CCRSs
and the X-ray-derived quantities to allow us to identify the
underlying mechanisms that link the AGN activity and the cooling
properties in clusters.

Shown in the left panel of Figure~\ref{fig:CorrPlots2} is the
bolometric X-ray cluster luminosity in the energy range 0.01$-$40~keV
as inferred from $\rosat$ and $\asca$ measurements
\citep{Reiprich2002}, $\lx$, versus the integrated radio luminosity
(see Section~\ref{IRL}) for the 48 clusters with CCRSs. For the SCC
clusters, shown as filled~(blue) circles, there is a clear positive
trend visible, although with a considerable spread. Since the X-ray
luminosity is related through scaling relations to other global
parameters of a cluster, such as the $\tvir$ and cluster mass, similar
correlations may be obtained between the radio power of a CCRS and
these quantities. This is the first time that the radio power of a
centrally located AGN, the prime candidate for counteracting the
cooling of the X-ray radiating ICM gas, has been shown to be
correlated with the large-scale cluster properties. This result
implies that there is a link between two regions, vastly differing in
scales; the region over which AGN accretion takes place, which is no
more than a few hundredth of a parsec, and the ICM, which extends out
to $1-2$ megaparsecs. Using the two-dimensional bisector linear
regression routine {\it BCES}, the trend between the $\lr$ and $\lx$
for SCC clusters may be quantified as below:
\begin{equation}
  \frac{\lr}{10^{42}~ h_{71}^{-2}~~\st{ergs~s^{-1}}} = \st a
  \times \,
  \left(\frac{\lx}{10^{44}~h_{71}^{-2}~\st{ergs~s^{-1}}}\right)^{\st b} \, ,  
\label{eqn5}
\end{equation}
where $\st a=0.03 \pm 0.01$ and $\st b= 1.38 \pm 0.16$. The Spearman
rank correlation coefficient of the fit is $0.64$ and the probability
of their being no correlation is $2.7 \times 10^{-4}$. The Pearson
correlation coefficient for the fit is larger and is equal to $\sim
 0.76$.

 Note that luminosity-luminosity plots should be considered with
 caution due to the common redshift-dependence in both the quantities
 \citep{Kembhavi1986,Akritas1996b,Merloni2006}.  Even though there are
 no censored data points~(upper limits) for the category of SCC
 clusters, in that every SCC has radio source at the center, spurious
 correlation may still be introduced due to the common dependence on
 the distance. In order to check for such an occurrence, we simulated
 randomized radio and X-ray luminosities confined to the observed
 ranges following the distributions, $n (\lx) \st{d}\lx \propto
 \lx^{-0.7}$ \citep{Boehringer2002} and $n (\lr) \st{d}\lr \propto
 \lr^{-0.78}$ \citep{Nagar2005}, where $n$ is the source number
 density. We assigned randomly distributed redshifts to the randomized
 luminosity data sets, according to the law $n \sim D^3_l$. These
 luminosities were re-observed after applying the X-ray flux limit,
 $f_{\textrm x}~(0.1-2.4)$~keV$ \ge 2 \times
 10^{-11}$~erg~s$^{-1}$~cm$^{-2}$, the same as that for the $\hiflux$
 sample, and the radio flux limit, $1.5$~mJy the average point-source
 detection limit for NVSS. The resulting Pearson correlation
 coefficients, $\rho_{\st P}$, were compared to the observed one.
 Based on these simulations, we compute the probability of having
 $\rho_{\st P} > 0.76$ and the correlation slope equal to or greater
 than that observed for a completely randomized set of X-ray and radio
 luminosities as less than $1\%$~(a spuriously induced correlation
 should produce a slope of around unity). This fraction increases to
 $\sim 2.5\%$ if instead the Spearman correlation coefficient is used
 and, if in addition, the observed probability of null hypothesis is
 used as a further constraint, i.e., the probability of null
 hypothesis for the simulated data sets should be lower than $2.8
 \times 10^{-4}$, then this fraction decreases to $\sim 1.5\%$.  Thus,
 we conclude that the probability of the observed correlation between
 the cluster X-ray luminosity and the radio luminosity of a CCRS to be
 spurious is very unlikely. However, that such an induced correlation
 is possible at a level of $\sim 3\%$ in the worst case scenario, is
 worth keeping in mind for past and future studies on similar topics.

Shown in the right panel of Figure~\ref{fig:CorrPlots2} is the radio
luminosity versus $\mdr$. This plot shows an even stronger trend than
that seen with $\lx$. This further strengthens the likelihood of a
coupling between gas cooling and the magnitude of the AGN activity.
The NCC clusters do not appear on this plot since these by definition
have no cooling radius, that is the central cooling time for these
clusters is greater than 7.7~Gyr, implying zero mass deposition rates.

\begin{figure}
  \centering
  \includegraphics[angle=-90, width=0.5\textwidth]{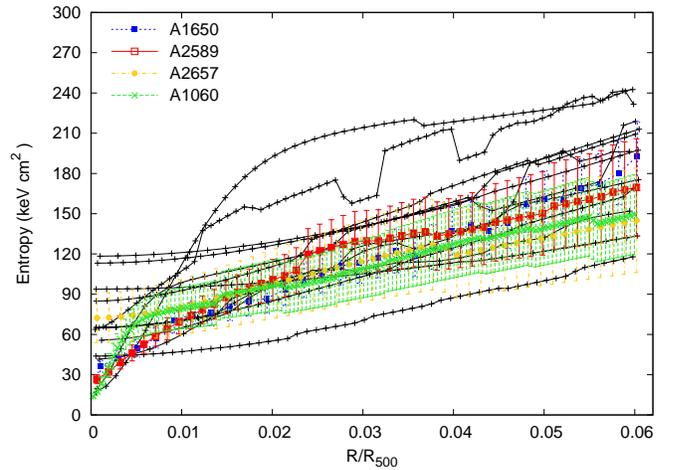}
  \caption{\small The entropy profiles of the WCC clusters. All with a
    central radio source are shown in black (plus) symbols and four
    without are shown in colored (non-plus) symbols. The errorbars on
    those with a central radio source are omitted for clarity. The
    strong jumps apparent in some of the entropy profiles are
    insignificant relative to the errorbars.}
  \label{fig:MDR_entropy}  
\end{figure}

There are two interesting subsets of clusters pertaining to the right
panel of Figure~\ref{fig:CorrPlots2}, which deserve attention. The
first subset comprises clusters which lack a CCRS but have $\mdr >
1~\mpy$, and the second subset, not shown in
Figure~\ref{fig:CorrPlots2}, comprises clusters which have a CCRS but
for which $\mdr = 0$. The former subset (denoted by black triangles in
Figure~\ref{fig:CorrPlots2}) consists of: A1650, A2589, A2657 and
A1060, with mass deposition rates ($93.7~\pm~28.2)~\mpy$,
($19.9~\pm~12.3)~\mpy$, ($15.4~\pm~10.9)~\mpy$ and
($6.6~\pm~3.0)~\mpy$ respectively. In order to understand the behavior
of the above four clusters, which are all WCC clusters, and to find
out whether there exists a quantity that separates them from the other
WCC clusters {\it with} a central radio source, we examined the
entropy profiles of these clusters. Entropy is a powerful tool which
provides information about two cluster parameters simultaneously - the
temperature and the density; $K(r) = kT(r)n(r)^{-2/3}$, where $r$ is
the radius from the cluster center.  Shown in
Figure~\ref{fig:MDR_entropy} are the entropy files of all but two WCC
clusters. The two exceptions are A3266 and A3667, which have no CCRS
but also have classical mass deposition rates consistent with zero. In
other words, these are cases at the border line between WCC and NCC
clusters and have, therefore, been excluded from
Figure~\ref{fig:MDR_entropy}. As can be seen, the entropy profiles of
these four clusters (shown as color curves with symbols other than
`$+$' symbols) are in no sense different from the rest. If anything,
the profiles of A1650, A2589 and A1060 seem to continue to fall with
decreasing clustercentric distance. This implies a steady increase in
the gas density with decreasing radius in these clusters and, hence,
relatively strong cooling. What is the source of heating in these
clusters?


\cite{Donahue2005} investigated one of the above radio-quiet CC
clusters, A1650, using $\chandra$ observations and proffered one of
the following two explanations for the absence of a radio AGN at the
cluster center; (1)~either the cluster has not reached the point where
heating is necessary, or (2)~the cluster experienced a major heating
event about 1~Gyr ago such that it has not required feedback since
then. Their conclusions are based on a lack of central temperature
gradient in A1650 and a markedly raised central entropy as compared to
other CC clusters with radio emission. Although the $\chandra$
observations used by us (including $\sim 200$~ks that became publicly
available in 2008) also imply an insignificant central temperature
drop ($T_0/\tvir \sim 0.8$), the estimated central entropy is not any
higher than the average central entropy of the rest of the WCC
clusters (Figure~\ref{fig:MDR_entropy}).

Although, all of the above four clusters pose a serious threat to the
AGN-regulated feedback fabric in cool-core clusters, A1650 is most
intriguing due to a high value of discrepancy between the expected and
measured mass deposition rates [$\mdr \sim (93.7 \pm 28.2)~\mpy$ and
$\smdr < 0.7~ \mpy$, where $\smdr$\footnote{A detailed description of
  how $\smdr$ is calculated can be found in \citet{Hudson2008}} is the
spectral mass deposition rate]. Interestingly, that there has been a
mention of a weak detection of a radio source at the center of A1650
by \citet{Dunn2006}, which in turn is based on the $\vla$ detection at
327~MHz by \cite{Markovic2004}, who give the total flux-density of the
radio source at this frequency as $59$~mJy. But we have been unable to
re-confirm this claim using the same observations as used by
\cite{Markovic2004} down to 3~mJy, three times the background noise.
As also pointed out by \cite{Donahue2005}, there is neither an
indication of a past AGN outburst, either in the form of
low-brightness diffuse lobe emission or cavities in the X-ray
emission, nor are there any signatures of a recent merger. This
cluster, along with its three companions, deserves further study in
order to analyze other possible sources of heating such as conduction,
intracluster supernovae or preheating.

The second subset corresponding to clusters with zero classical mass
deposition rates but which contain a CCRS consists of: A3391, A3395s,
A3376, A0400, A1656, A3158, A2147, MKW8 (in decreasing order of
$\lr$). These are NCC by definition and show signs of cluster mergers
at different stages, as do most of the other NCC clusters in our
sample \citep{Hudson2008}, based either on their X-ray
properties~(presence of subclumps or non-negligible separation between
the X-ray emission peak and emission weighted centroid) or radio
properties~(presence of radio halos or relics or both). But contrary
to the clusters in this subset, the remaining NCC clusters are devoid
of a central radio source, supporting the idea that the presence of
AGN is tightly correlated to gas cooling in clusters. The obvious
question that then surfaces is, how does this subset fit into the
AGN-heating and gas-cooling machinery? There are several solutions to
this apparent contradiction.  Firstly, a well-known fact $-$ AGN are
found at all locations in clusters and not only at the cluster centers
\citep[although with an increasing probability with decreasing
clustercentric distance, see][]{Morrison2003,Best2007}. There also
exist field-AGN with no apparent reservoir of bulk cool gas, such as
is available in clusters.  Hence, there evidently are mechanisms other
than those related to cluster cooling that can trigger radio nuclear
activity in galaxies.  Secondly, there is evidence that mergers may
play a role in activating the central engines of AGN by transferring
gas to the cluster galaxies and providing material for both, AGN
accretion and also star formation \citep{Owen1999}. But alongside
there also are contradictory findings according to which mergers may
as well strip away gas from galaxies and result in inhibition of both
the processes \citep[e.g.~see][]{Gia2004}. Thirdly, a configuration
containing a NCC cluster with zero mass deposition rate and a central
radio source may be obtained through a merger between a NCC cluster
and a CC cluster, latter harboring a central AGN, in such a way that
it results in disruption of the cool core and leaves behind only
traces of the past cooling activity. The simulations of
\cite{Burns2008} indeed show that NCC clusters are formed as a result
of major mergers right at the beginning of their evolution, whereby
they grow in time at the expense of CC clusters. As an example, A2634
is a WCC cluster based on the central cooling time but in most other
respects is closer to being a NCC cluster than a WCC
cluster. According to \cite{Hudson2008}, the X-ray morphology of this
cluster is consistent with that of a merging cluster. Yet it has a
cool core (short $\ct$), and so A2634 might be a strong candidate for
such a case where the cool core is being destroyed by a major merger.

Excluding the cluster on the upper-left corner~(A2634) of
Figure~\ref{fig:CorrPlots2} and the aforementioned subsets of
outliers, the powerlaw fit for SCC and WCC clusters using the {\it
  BCES} routine gives,
\begin{equation}
  \frac{\lr}{10^{42}~ h_{71}^{-2}~~\st{ergs~s^{-1}}} = \st a
  \times \, \left(\frac{\mdr}{\ms~\st{yr}^{-1}}\right)^{\st b} \, ,
\label{eqn6}  
\end{equation} 
where $\st a = (4.7 \pm 5.0) \times 10^{-5}$ and $\st b = 1.69 \pm
0.25$. It can be argued that this exercise may be more appropriate if
we use the spectrally determined mass deposition rate, $\smdr$, as
opposed to the classical one, $\mdr$, as the former gives the actual
observed rate at which the gas is cooling out and accreting onto the
supermassive black hole.  $\smdr$ is, in fact, the fuel for the
central AGN and should be correlated with the AGN output. On the other
hand, the question that we are trying to address here is whether the
AGN output can account for the difference between the
classical~(predicted) and spectral~(observed) mass deposition rate
and, thereby, provide a solution to the cooling flow puzzle. In
addition, excluding clusters for which $\smdr$ is consistent with
zero, the classical mass deposition rates exceed the spectral mass
deposition rates on average by a factor of 15. Hence, $(\mdr-\smdr)
\sim \mdr$ and the correlation between the radio luminosity and the
two types of mass deposition rates is expected to be similar. We also
point out that the spectral mass deposition rates estimated using the
ACIS instrument on $\chandra$ are, due to its low spectral resolving
power, only moderately accurate. Therefore, any cross-correlation with
the spectral mass deposition rates will have a large uncertainty
making robust interpretations difficult.

Finally, we interpret the strong correlation seen between $\mdr$ and
$\lr$ as supporting evidence for a feedback system in which the AGN
activity is more enhanced in clusters with higher mass deposition
rates, and the AGN in turn quenches the cooling of gas by heating the
ambient medium. We note that a similar result was obtained by
\citet{Peres1998} between $5$~GHz spectral radio luminosity and the
mass deposition rate determined with $\rosat$, although for a much
smaller subset constituting only 15 galaxy clusters of the B55 sample
\citep{Edge1990}.

\subsection{Brightest cluster galaxies~(BCGs)}
\label{bcg}

The brightest cluster galaxies are unique in terms of their
high-luminosity and proximity to the centers of their host clusters.
The BCGs are extremely interesting objects and have long been subjects
of a wide range of studies. At one extreme, their formation and
evolution is closely tied with the Mpc scale cluster environment in
which they reside. At the other extreme in the hierarchy of structure
formation, BCGs are just one level above the $\sim 10^{-4}$~pc scale
supermassive black holes~(SMBH). The BCG bulge properties, such as the
optical bulge luminosity and stellar velocity dispersion, obey certain
scaling relations that permit indirect estimation of the mass of SMBHs
\citep[e.g.][]{Kormendy1995,Ferrarese2000}.

In this section, we correlate the BCG magnitudes or, equivalently, the
mass of the SMBHs with the AGN radio luminosity and the large-scale
X-ray properties of galaxy clusters. The mass of the SMBH, $\mbh$, is
derived using the scaling relation between the near-infrared~(NIR)
bulge magnitude and the inferred $\mbh$, as deduced by
\cite{Marconi2003}. Further, we test whether there are any
distinctions in the BCGs properties amongst the three different types
of clusters (SCC, WCC and NCC clusters).

The BCG apparent magnitudes were taken from the Two Micron All-Sky
Survey \citep[2MASS][]{Skrut2006} Extended Source Catalog
(XSC\footnote{http://irsa.ipac.caltech.edu/Missions/2mass.html}). We
used the 2MASS total magnitudes in $K$-band~($2.16~\mu$m),
$km_{\st{ext}}$, estimated from extrapolation of the surface
brightness profiles~(SBP). The extended source detection limit for
$K$-band at 10$-\sigma$ is 13.5~mag and the uncertainties range from
0.02$-$0.23~mag with a mean of 0.06~mag. In short, the 2MASS SBPs have
been derived from fitting a modified Sersic function to the elliptical
radial light distribution of the BCGs. The total magnitudes are
estimated from summing two terms. The first term corresponds to the
isophotal magnitude estimated from fitting an ellipse to the standard
isophote of mean surface brightness of $K_{20} =
20$~mag~arcsec$^{-2}$. The second term is derived by integrating the
best-fitting Sersic law starting from the standard isophote, $r_{20}$,
out to a delimiting isophote, which is typically about four scale
lengths. This 2MASS strategy insures that the total flux of a galaxy
is recovered.

The BCGs in 2MASS were located by searching around the brightest
cluster galaxy in each cluster using an initial compilation, kindly
provided by Heinz Andernach, based on the available data at NASA/IPAC
Extragalactic Database (NED\footnote{http://nedwww.ipac.caltech.edu})
and Hyperleda\footnote{http://leda-univ-lyon1.fr}. All the BCGs were
found within $5^{''}$ of the given search position except in the case
of three clusters, A2204~(SCC), A2065~(WCC) and A2163~(NCC), where,
based on visual inspection, 2MASS did not manage to locate the {\it
  right} galaxies (i.e. even though within $5^{''}$ of the given
search position, the putative BCGs were not found in the 2MASS-XSC
catalog). For these three clusters, we retrieved the 2MASS $K$-band
Atlas images which have a plate scale of $1^{''}/$pix, and fitted the
BCG surface brightness distributions using the two-dimensional
galaxy-fitting program, GALFIT \citep{Peng2002}. In order to be
consistent with the 2MASS fitting routines, we restricted the
functional form to Sersic models. While the SBPs of the BCGS of A2065
and A2163 were well-reproduced with a single-component Sersic models,
the BCG in A2204 required a double-component Sersic model. For
comparisons of the $K$-band magnitudes between the 2MASS-XSC and
GALFIT estimates, we carried out tests by applying GALFIT to a few of
the BCGs present in the XSC catalog and, hence, with known 2MASS
magnitudes. We found that whereas GALFIT systematically underestimates
the magnitude for bright BCGs ($km_{\st{ext}} < 10$~mag) by about
$10\%$, the GALFIT magnitudes are consistent with the 2MASS-XSC
magnitudes for faint BCGs ($km_{\st{ext}} > 10$~mag) to within
$3\%$. Since the GALFIT magnitudes of the BCGs in A2204, A2065 and
A2163 are all fainter than 12~mag (but brighter than the detection
limit of 13.5~mag), we deem them to be trustworthy to within $5\%$.

The apparent magnitudes from 2MASS were corrected for Galactic
extinction using values from \citet{Schlegel1998}, the typical
correction values being small $-$ on the order of $\sim 10^{-2}$, and
were then converted into absolute magnitudes using the redshifts
compiled from NED. We did not apply any $k-$ correction since these
galaxies are all nearby.

\subsubsection{Supermassive black hole mass and radio luminosity of
  the BCG}
\label{bcg2}

There has been a lot of debate over the use of BCG scaling relations
for determinations of $\mbh$. This is due to the fact that many BCGs
are often accompanied by low surface brightness envelopes extending
out to, as far as, several hundred kiloparsecs \citep{Gonzalez2005}.
These are the well-known ``cD galaxies''. The extended envelopes, also
known as intracluster light~(ICL) \citep{Lin2004}, are thought to
either represent debris accumulated over the merger history of the BCG
or from tidal stripping of other cluster galaxies. The extended
emission may as well originate from stars forming out of the condensed
gas in cooling clusters. An important investigation we will follow-up
in a subsequent paper is to study how the luminosity of these
envelopes correlates with cooling parameters, such as the $\ct$,
$\mdr$, $\smdr$ etc. This requires careful decomposition of the BCG
light profile into an inner component, associated just with the
galaxy, and an outer flatter component representing the
ICL. \cite{Seigar2007} fitted analytical models with two Sersic
components to separately measure the profiles of the central and
extended parts of 5 cD galaxies and showed that the contribution of
the envelopes to the total light is around 60\% to 80\%.

\begin{figure}
  \centering
  \includegraphics[angle=-90, width=0.5\textwidth]{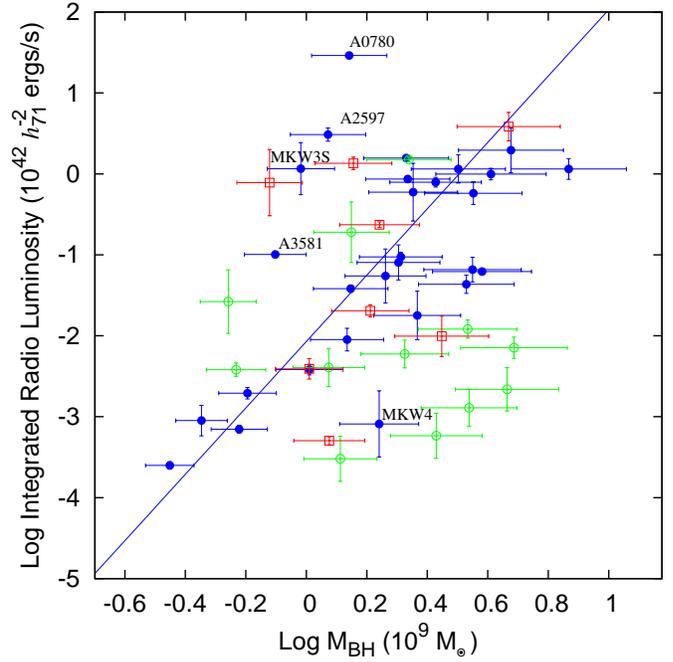}
  \caption{\small The mass of the SMBH versus the total radio
    luminosity of the BCG in SCC (blue filled circles), WCC (green
    open circles) and NCC (red open squares) clusters. Visible is a
    weak trend given by solid blue line for the SCC clusters such that
    the $\lr$ of the BCG increases with the SMBH mass. Labeled are the
    five SCC outliers.}
  \label{fig:Mbh_rlum}  
\end{figure}

Despite the above factors, recent studies
\citep{Batcheldor2007,Marconi2003} have shown that the SMBH masses
derived from the scaling relation using the NIR bulge magnitudes of
the BCGs are consistent with those derived from stellar velocity
dispersions ($\sigma_{\st v}$), both the relations yielding similar
amount of scatter. This is not quite true for $V$- or $B$-band
magnitudes though, where the scatter is much higher compared to the
preferred $\sigma_{\st v}$-$\mbh$ relation. \cite{Batcheldor2007}
attribute the scatter to inclusion of luminosity from the
outer-envelopes of the cD galaxies, which may be more pronounced in
the $V$- or $B$- band, especially if the outer-envelope light
represents on-going star-formation, and which might not have anything
to do with the central galaxy dynamics \citep[however, see][for a
difference of opinion]{Lauer2007}. Based on the studies on the NIR
bulge magnitude-$\mbh$ relation, we use the following scaling relation
\citep{Marconi2003} to derive the SMBH masses :

\begin{equation}
\st{log}_{10}\left({\frac{\mbh}{M_{\odot}}}\right) = \st a + \st b 
\left[\st{log}_{10}\left({\frac{\lbcg}{L_{\odot}}} \right)
- 10.9 \right]  \; , 
\label{lbcg-mbh}
\end{equation}
where $\st a=8.21 \pm 0.07$ and $\st b=1.13 \pm 0.12$. In order to
convert the absolute magnitudes into luminosities in units of $K$-band
solar luminosity, we used the absolute $K$-band solar magnitude equal
to 3.32~mag \citep{Colina1997}. In Figure~\ref{fig:Mbh_rlum} we
present the SMBH masses versus the $\lr$ for the 48 CCRSs (BCGs) in
our sample. Whereas on the whole there appears to be a poor
correlation of increasing AGN radio output with increasing $\mbh$,
categorization of clusters based on $\ct$ results in a trend to appear
between the two quantities but only for the SCC clusters. The best-fit
powerlaw using the two-dimensional fitting algorithm $BCES$ for the
SCC clusters is:
\begin{equation}
  \frac{\lr}{10^{42}~ h_{71}^{-2}~~\st{ergs~s^{-1}}} =
  (0.008 \pm 0.004) \times \, \left(\frac{\mbh}{10^9~\ms}\right)^{4.10
    \pm 0.42}  \, .
  \label{eqn7}  
\end{equation} 
The Spearman rank correlation coefficient of the fit is $0.46$ and the
probability of their being no correlation is $1.3 \times 10^{-2}$. The
Pearson correlation analysis yields a larger correlation coefficient
($0.59$) and a much smaller probability of the null-hypothesis ($9
\times 10^{-4}$).

\begin{figure*}
  \centering
  \includegraphics[angle=-90,width=\textwidth]{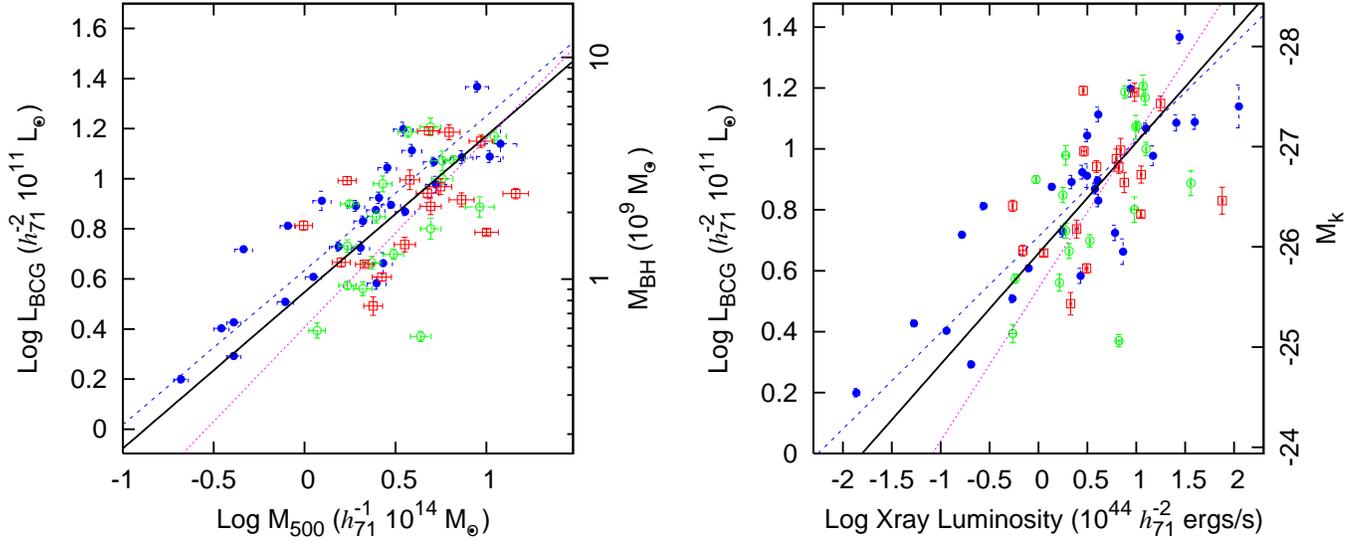}
  \caption{\small The 2MASS $K$-band total magnitudes of the 64 BCGs
    in the $\hiflux$ sample versus $\mvir$ in the left panel and
    $\rosat$ bolometric cluster X-ray luminosity in the right panel.
    The different symbols and colors denote SCC (filled blue circles)
    clusters, WCC (non-filled green circles) clusters and NCC
    (non-filled red squares) clusters. The solid~(black), the dashed
    (blue) and the dotted~(pink) lines denote the best-fit powerlaws
    for all clusters, only SCC clusters, and only non-SCC (WCCs+NCCs)
    respectively.}
  \label{fig:Mag}  
\end{figure*}    

Several studies have explored the correlation between the SMBH mass
and radio luminosity of the embedded AGN, leading to contradictory
results. Whereas it is clear that the radio loudness, the ratio of the
radio to the optical luminosity of an AGN, is a strong function of the
black hole mass \citep{Laor2000}, whether or not there exists a tight
relation between $\lr$ and $\mbh$ is still a matter of
debate. \citet{Franceschini1998} derived a tight relation between the
radio luminosity at 5~GHz, $L_{\st{5~GHz}}$, and $\mbh$ using a sample
of 13 nearby early-type weakly active galaxies, such that
$L_{\st{5~GHz}} \propto \mbh^{2.66}$. \citet{Lacy2001} also find a
similar correlation for a sample of steep-spectrum radio sources,
albeit leading to a flatter slope of 1.4. Although \citet{Laor2000}
confirmed this trend of increasing radio luminosity with increasing
black hole mass using a much larger sample of 29 nearby galaxies and
89 PG (Palomar-Green) quasars, the trend they obtain is weak and
presents a lot of scatter. As an example of dissenting views,
\citet{Liu2006} find no relation of $L_{\st{5~GHz}}$ against the black
hole mass. On the other hand, they find a strong correlation between
the jet power and the black hole mass, where they used the flux
density of the radio lobes at 151~MHz as a proxy to determine the jet
power. Based on the above results it is not yet clear how the radio
luminosity of an AGN scales with the black hole mass. It may be that
different black hole efficiencies, conversion rates from the total AGN
power to the radiative power of the jet and large-scale environmental
effects from source to source, cause the radio luminosity to display a
wide range of values for a given black hole mass.

In pursuit of determining whether or not there is an underlying
physical mechanism that ties the radio power to the mass of the SMBH,
Figure~\ref{fig:Mbh_rlum} presents an interesting outlook. We
investigate for the first time the dependence of $\lr$ on $\mbh$
taking into consideration the different environments (SCC, WCC and NCC
clusters) in which the centrally located radio sources reside.
According to the simulations of \citet{Burns2008}, CC clusters grow an
enhanced cool-core gradually and steadily via small mergers, unlike
NCC clusters which experience major mergers early in their history.
Hence, repetitive small mergers in addition to providing material for
the growth of the cool core in CC clusters, may also result in a
quasi-steady mass accretion rate onto the SMBH in CC clusters. The
fact that only the CCRSs in SCC clusters show some level of
correlation between $\mbh$ and $\lr$ implies that the AGN output in
these systems is proportional to the average mass-accretion rate onto
the black hole, thereby, balancing the radiative losses.


At the time of writing this article, MKW4, an outlier below the
best-fit line in Figure~\ref{fig:Mbh_rlum}, had no low-frequency
($<1.4$~GHz) radio data available. This inspired us to acquire 327~MHz
$\vla$ data for this cluster, which is work-in-progress.  MKW4 is an
interesting cluster in that high-frequency (1.4~GHz and 5~GHz) $\vla$
archival data showed a cluster of closely spaced point-sources
$\gtrsim1$~arcmin to the NE of the BCG radio emission.  Furthermore,
radio emission from the BCG was only detected at 5~GHz and not at
1.4~GHz, the latter having had only 2-minute on-source integration
time. The preliminary 327~MHz image of MKW4 revealed the same feature
NE of the BCG but due to insufficient resolution, it still remained
unclear whether this emission was associated with the nucleus of the
CCRS or corresponded to a high-redshift cluster system; the latter
conclusion being based on the proximity of these point sources to each
other. Assuming that the emission in the NE reflected the region of
intense interaction between the lobe and ICM and, hence, was a part of
the central radio source, would have caused the $\lr$ to increase 30
fold and for it to no longer be an outlier. To confirm this
hypothesis, we acquired dynamic $\vla$ time at 1.4~GHz in CnD
configuration with 3~hrs integration time. Even though the new 1.4~GHz
data show no signature of a connection between the CCRS and the bunch
point-source emission in the NE, it clearly shows emission from the
CCRS, which went undetected with the old archival data. This itself
increased the total radio output by more than an order of magnitude
bringing this cluster closer towards the observed $\mbh$-$\lr$ trend
for the SCC clusters (this change is incorporated in
Figure~\ref{fig:Mbh_rlum}). The follow-up study of the CCRS in MKW4
underlines the importance of obtaining reliable spectra of radio
sources.

In addition to MKW4, there are four other outliers in
Figure~\ref{fig:Mbh_rlum} which lie above the best-fit relation:
A0780, A2597, MKW3S and A3581. We believe these systems might be ones
which have experienced powerful radio outbursts in recent past
resulting in the present AGN heating rate to be greater than the
average rate at which the mass has been accreting onto the SMBH since
the formation of the cool core. Similarly, MKW4 may reflect a CCRS
that has been caught just at the beginning of another heating cycle
and has yet to reach its peak radio activity. These outliers, both
below and above the best-fit line may be reflective of the episodic
nature of CCRSs in some systems. Another plausible reason for these
outliers may be the weakness in the underlying assumption of the
integrated luminosity of the BCG as a robust indicator of the bulge
luminosity in all the cases.

\subsubsection{Large-scale cluster properties and BCGs}
\label{bcg1}

Inherent scalings between BCGs and clusters have been implied in
numerous observational studies \citep[e.g.][]{Lin2004,Brough2008} as
well as cosmological simulations \citep{Zheng2007,Cooray2005}. In this
context, formation and evolution of BCGs and its dependence on the
host cluster is an important tool to understand these scalings
observed between the BCG luminosity and the host halo mass and X-ray
luminosity. There are several proposed BCG evolution scenarios to
support these observations, such as (1)~dynamical-friction governed
galactic cannibalism, (2) rapid mergers between galaxies during the
epoch of cluster formation, (3)~co-evolution of BCGs with cluster
growth due to mergers embracing the paradigm of hierarchical structure
formation and (4)~cooling-flows. While the first two scenarios may
contribute significantly during the early epoch of BCG formation,
\citet{Lin2004} and \citet{Brough2008} argue that the BCGs co-evolve
with the host clusters via mergers with the BCGs of the falling
subclusters, which lead to subsequent growth of the BCG luminosity
with increasing cluster mass.

We present in this section the BCG-host cluster correlations for the
$\hiflux$ sample, with the aim to extend previous studies by analysing
a low-z flux-limited sample unique in its completeness and homogeneity
in the ways of obtaining the X-ray and NIR quantities. The following
study is unique in an additional aspect in that we bear in mind the
possibility of different growth histories for the BCGs corresponding
to CC and NCC clusters.

Shown in Figure~\ref{fig:Mag} is $\mvir$ versus $\lbcg$ in the left
panel and $\lx$ versus $\lbcg$ in the right panel. Shown in different
symbols and colors are the three different types of clusters. Whereas
both the panels clearly indicate that the BCG grows with the cluster
size, the left panel additionally shows a segregation between the SCC
clusters (blue filled circles) and the non-SCC clusters. A possible
reason for such a separation may be related to the continuous growth
of BCGs, as already mentioned in Section~\ref{bcg2}, due to iterative
small-scale mergers through which the CC clusters grow. The BCGs in
NCCs, on the other hand, form constituents of non-relaxed cluster
environments with a history of one or more major mergers, and the
subsequent heating at the central regions may hinder further BCG
growth at the same rate as that of BCGs in CC clusters. This argument
is supported by the fact that on comparing the radial profiles of
numerically simulated CC and NCC clusters, \citet{Burns2008} observe
an excess of baryons in CC clusters relative to NCC clusters.

An interesting note is that this segregation appears only between
$\lbcg$ versus $\mvir$ and not between $\lbcg$ versus $\lx$. However,
a similar intrinsic separation for CC and NCC clusters is seen between
$\tvir$ and $\lx$. This is attributed to the fact that at a given
temperature, SCC clusters have a higher luminosity as compared to
non-SCC clusters due to an increased gas density at the center. This
causes the SCC clusters to form an envelope towards the higher
luminosity end. Since $\mvir$ has been derived from $\tvir$, it may be
that the resulting magnitude of separation between $\mvir$ and $\lx$
cancels with that between $\lbcg$ and $\mvir$.

The estimated best-fit powerlaw for the $\lbcg$-$\mvir$ relation based
on the {\it BCES} algorithm is:
\begin{equation}
  \frac{\lbcg}{10^{11}~h_{71}^{-2}~ \ls} = \st a \times 
  \left(\frac{\mvir}{10^{14}~\ms~\st{yr}^{-1}}\right)^{\st b} 
\label{eqn8}  
\end{equation} 
where $\st a = 3.525 \pm 0.277$ and $\st b = 0.624 \pm 0.054$ for all
clusters, $\st a = 4.305 \pm 0.290$ and $\st b = 0.616 \pm 0.005$ for
the SCC clusters only and $\st a = 2.552 \pm 0.362$ and $\st b = 0.752
\pm 0.095$ for the non-SCC clusters~(CC clusters) only. The Spearman
rank correlation coefficients for the fits are $0.67$, $0.87$ and
$0.62$ for all, SCC and non-SCC clusters, respectively. Looking at the
fit results for all clusters and SCC clusters only, it is clearly seen
that the SCC clusters have a higher normalization by about 20\%(since
the two fits have the same slope within the 1-$\sigma$ errorbars, the
normalizations may be directly compared).  The slope of the obtained
$\lbcg$-$\mvir$ relation is steeper than the values derived in other
works, which tend to center around $0.3$, even though presenting a
wide range from $0.1$ to $0.5$
\citep{Lin2004,Brough2008,Popesso2007,Hansen2007}.

Going a step further, we may combine Equations~(\ref{lbcg-mbh}) and
(\ref{eqn8}) to derive a relation between the SMBH mass and the
cluster mass:
\begin{equation}
  \frac{\mbh}{10^9~\ms} = \st a \times
  \left(\frac{\mvir}{10^{14}~\ms~\st{yr}^{-1}}\right)^{\st b} 
\label{eqn9}
\end{equation}
where $\st a = 0.98 \pm 0.08$ and $\st b = 0.61 \pm 0.06$ for all
clusters, $\st a = 1.15 \pm 0.09$ and $\st b = 0.63 \pm 0.06$ for the
SCC clusters only and $\st a = 0.75 \pm 0.14$ and $\st b = 0.70 \pm
0.12$ for the non-SCC clusters. Even though indirectly derived using
the BCG bulge luminosities, such a correlation could be indicative of
a fundamental relation between the host cluster halo and central SMBH
similar to the galaxy bulge mass-black hole mass relation.

Similarly, the best-fit powerlaw for the $\lbcg$-$\lx$ relation is
\begin{equation}
  \frac{\lbcg}{10^{11}~h_{71}^{-2}~ \ls}
  = \st a \times
  \left(\frac{\lx}{10^{44}~h_{71}^{-2}~~\st{ergs~ s}^{-1}}  
  \right)^{\st b}     
\label{eqn10}  
\end{equation} 
where $\st a = 4.54 \pm 0.34$ and $\st b = 0.36 \pm 0.03$ for all
clusters, $\st a = 5.15 \pm 0.38$ and $\st b = 0.32 \pm 0.03$ for the
SCC clusters only and $\st a = 3.49 \pm 5.09$ and $\st b = 0.50 \pm
0.07$ for the non-SCC clusters.  The Spearman rank correlation
coefficients for the fits are $0.68$, $0.82$ and $0.55$ for all, SCC
and the non-SCC clusters, respectively. The $\lbcg$-$\lx$ and
$\lbcg$-$\mvir$ relations show that there is a statistically
significant difference in slopes between SCC and non-SCC clusters.
Similar correlations were obtained by \cite{Katayama2003} using a
larger sample, which also contains the $\hiflux$ sample.  However,
their results show much higher scatter between the BCG luminosity-host
halo mass, which is due likely to the use of optical B-band magnitudes
which have larger errorbars of mean value $\sim 0.2$~mag. Further,
\cite{Katayama2003} correlated the optical magnitude versus the total
mass of the host cluster defined as the integrated mass within a fixed
metric radius of 5~Mpc, for which both, statistical as well as
systematic uncertainties are expected to be larger.

The above results, which highlight a strong dependence of the BCG NIR
magnitudes on the scale of the host cluster, however, are at variance
with those obtained by \cite{Brough2002}. They claim that any division
between BCGs in low-$\lx$ and high-$\lx$ clusters as seen in high-$z$
clusters, disappears for clusters with $z\leq 0.1$.  The input low-$z$
sample ($z \leq 0.1$) studied by \cite{Brough2002} consisted of 150
Abell clusters with a flux limit of $f_{\textrm x}~(0.1-2.4)$~keV$ = 3
\times 10^{-12}$~erg~s$^{-1}$~cm$^{-2}$ in the $\rosat$ {\it hard}
band (0.5$-$2.0~keV). After matching these with the 2MASS catalog
resulted in a final sample comprising 76 clusters with only those BCGs
which have robust 2MASS magnitude measurements in the $K$-band. We
argue that the contradiction in results might stem from differences in
the aperture radii used, within which the magnitudes are calculated.
The 2MASS database provides galaxy magnitudes based on a suite of
apertures. While \cite{Brough2002} employ integrated magnitudes
measured using circular apertures of a fixed metric radius of
12.5$\hinv{-1}$kpc, we use, as explained above, the total aperture
radii by extrapolating the SBPs, the mean of which for the BCGs in our
study is $\sim 50\hinv{-1}$kpc. From this we conclude that the total
BCG magnitude is a better quantity to use, as opposed to the BCG core
magnitude, for detecting correlations with global cluster properties.

\section{Discussion}
\label{discussion}

The riddle of cooling-flows in clusters has continued to baffle us.
Recent high-resolution $\chandra$ images revealed radio-loud AGN
embedded in the centers of cool-core~(CC) clusters surrounded by
regions emptied of the X-ray emitting gas, suggesting a strong tie
between the cluster central radio source and the cooling of ICM. Since
the discovery of numerous AGN-blown bubbles in the atmospheres of CC
clusters, various modes of energy transfer from the AGN to the ICM
have been investigated. While the most successful mode has turned out
to be dissipation of energy stored in the radio bubbles as they rise
buoyantly through the ICM, previous studies have fallen short of a
thorough investigation of the relation between the radio luminosity of
the centrally located AGN and cooling properties of a CC cluster.

It is now a widely accepted fact that the AGN activity is triggered by
gas accretion onto the central black hole. Playing devil's advocate,
it may then be argued that the AGN output is only to be expected to
scale with the mass accretion rate. Under the assumption that the cool
gas flows from the outer cluster regions to the very centers of the
BCGs and serves as the fuel for the black hole, it is not surprising
that the radio luminosity of an AGN should scale with the cluster
mass, and also, even though to a lesser degree, the inverse of the
cooling time of the gas. This explains the underlying trend seen in
Figure~\ref{fig:CorrPlots2} between the $\mdr$ and the radio output of
CCRSs in CC clusters. However, the picture thus developed so far does
not contain any ingredients reflecting on a self-regulated cycle
formed between gas cooling, star formation and AGN heating. In other
words, it may well be that even though AGN activity enhances with
cluster scale, the cooling of ICM is regulated by an altogether
different process, such as cluster mergers.

The first strong argument in favor of AGN-regulated heating comes from
the observation that the AGN fraction increases with decreasing
central cooling time, $\ct$ being the best diagnostic to distinguish
CC from NCC clusters. The study by \cite{Rafferty2008} shows that the
central star-formation rate also is a strong function of $\ct$ (see
Section~\ref{fractions}). That only clusters with short $\ct < 1$~Gyr
have increasing on-going star formation with decreasing clustercentric
distance, implies a chain of intricately linked processes which
maintain heating and cooling rates in cluster atmospheres. These
results together call for a feedback process in which AGN heating
becomes more of a requisite in clusters with shorter cooling times.
This may either be in form of huge AGN outbursts which heat the
surrounding cluster gas, the effect of which lasts for several cooling
cycles (such as Hydra-A, MKW3S, A2597 and A3581), or in form of
short-lived AGN outbursts which are repeated after short intervals. A
recent study by \cite{Shabala2008} has shown that the CCRSs in more
massive clusters undergo AGN outbursts more frequently than the AGN in
their less massive counterparts. Additionally, radio source models
employed by \cite{Shabala2008} show that the duration of the on-state
of an AGN has the same relation with the stellar mass as the mass
deposition rate has with the stellar mass \citep{Best2005}, suggesting
the switching on and off of an AGN resulting directly from either
availability or depletion of cool cluster gas.

More recently, \cite{Voit2008} have provided evidence that the AGN
activity, ICM cooling and star-formation might all be linked together
through the process of electron thermal conduction. According to their
study, the efficacy of thermal conduction depends on the size of the
temperature inhomogeneities relative to the critical length scale
associated with conduction, $\lambda_{\st f}$. The state of
equilibrium between radiation losses and conduction gain can be
equivalently expressed in terms of $\lambda_{\st f}$ and $K$. Based on
above arguments, conduction sets an entropy threshold such that only
those clusters whose central entropy is less than $30$~keV~cm$^{-2}$
show star formation, in the form of H$\alpha$, and enhanced AGN
activity. Above this threshold, conduction is a viable heating
mechanism. Further, since gas entropy is very closely linked to the
central cooling time through the relation, $\ct \propto K^{3/2}/T$,
such that 0.6~Gyr~(1~Gyr) corresponds approximately to
$30$~keV~cm$^{-2}$~($43$~keV~cm$^{-2}$), these observations are also
in concordance with our results on the central temperature drop and
cuspiness displayed in Figure~\ref{fig:1Gyr}. The central entropy (or
central cooling time) threshold may be an explanation for observing an
abrupt central temperature drop and an increase in cuspiness for
clusters with cooling times shorter than 1~Gyr. In the cluster regime
with $\ct < 1$~Gyr, AGN heating is the dominant balancing mechanism to
cooling.

We point out that there is non-negligible scatter in
Figures~\ref{fig:CorrPlots2} and \ref{fig:Mbh_rlum}, the origin of
which could be either extrinsic or intrinsic. In the presence of an
AGN-regulated feedback, an intrinsic scatter may imply that the
synchrotron luminosity, which is only a small fraction of the total
AGN output, is not a very reliable quantity to use to establish the
AGN-ICM interaction. It has been noted in previous studies, that the
ratio of kinetic to radiative AGN power indeed shows a broad range,
from a few to a several thousands
\citep[e.g.][]{Birzan2008,Birzan2004}. Kinetic AGN luminosity may be a
more robust measure of the total AGN feedback. For this, one requires
the radio morphology of jets and lobes overlaid on X-ray images to
help find or confirm the X-ray cavities. The census of X-ray cavities
is highly incomplete since they are of very low contrast, yet they are
important contributions to the heating budget.  An extrinsic scatter
in the plots would point at inaccurate measurements of observable
parameters at both the wavebands, radio~(incomplete spectral
information) as well as X-ray~(imprecise mass deposition rates due to
spectral resolution power of $\it{ACIS}$ on $\chandra$).

However, strong correlations found in this work between the total AGN
radio power and various cluster parameters lend confidence in
synchrotron luminosity as a fairly good measure of the cooling
activity in clusters. These correlations also provide us with
motivation to continue our work to acquire low-frequency radio
measurements for CCRSs which have no reliable data below 500~MHz
(constituting 35\% of the $\hiflux$ sample, see Section~\ref{IRL}).
To achieve this goal, we are awaiting proprietary data for all but two
of the remaining $35\%$ clusters with $\vla$ at 325~GHz and with
$\gmrt$ at dual-frequency band 610~MHz/235~MHz.

\section{Conclusions}
\label{conclusions}

We have presented a detailed joint analysis of the brightest complete
local sample of galaxy clusters, $\hiflux$, using high-resolution
X-ray data acquired from $\chandra$ and radio data compiled from
various sources spanning a wide range of frequencies. This study was
conducted so as to explore the role of AGN in the centers of galaxy
clusters accompanied with the cooling flow problem. The main results
of this study are:

\begin{enumerate}
\item We find that the integrated radio luminosity~($\lr$) of a
  cluster central radio source is tightly correlated to its 1.4~GHz
  monochromatic luminosity (see Figure~\ref{fig:RadSpecLum}) with the
  exception of a few outliers. This correlation is quantified as $\lr
  \propto L^{0.98 \pm 0.01}_{1.4~\st{\small GHz}}$
  (Equation~\ref{eqn1}). To estimate the integrated radio luminosity
  of a CCRS, a special effort was made to compile low-frequency radio
  measurements in order to get an accurate measure of the total
  radiative output of the centrally located AGN.
\item The best property to diagnose a cool-core cluster with high
  quality data is the central cooling time, $\ct$. Based on $\ct$,
  there is an increasing probability for the brightest cluster
  galaxy~(BCG) closest to the X-ray peak emission to harbor an AGN
  with decreasing cooling time (Figure~\ref{fig:Fraction}). The
  percentage of AGN in three bins ordered in cooling time, strong cool
  core~(SCC, $\ct \le 1$~Gyr), weak cool-core~(WCC, $1$~Gyr $\le \ct
  \le 7.7$~Gyr) and non-cool-core~(NCC, $\ct \ge 7.7$~Gyr) clusters,
  is $100\%$, $67\%$ and $45\%$ respectively.
\item There is a trend between the $\lr$ and $\ct$ such that the
  former increases with decreasing cooling time. This is shown in
  Figure~\ref{fig:CorrPlots1}, although with a large scatter
  especially towards short $\ct$ where the trend appears to break
  down.
\item The total radio output of a CCRS scales with the cluster
  size~(e.g.~X-ray luminosity). This correlation is particularly
  noticeable in SCC clusters (see Figure~\ref{fig:CorrPlots2}, left
  panel). The best-fit powerlaw for the SCC clusters is $\lr \propto
  \lx^{1.38 \pm 0.16}$ (Equation~\ref{eqn5}).
\item The total radio output of cool-core clusters~(SCC and WCC
  clusters) shows a tight correlation with the classical mass
  deposition rate, $\mdr$ (Figure~\ref{fig:CorrPlots2}, right
  panel). The correlation is given by $\lr \propto \mdr^{1.69 \pm
    0.25}$ (Equation~\ref{eqn6}).
\item The radio luminosity of the central radio source shows a weak
  trend with the mass of the supermassive black hole, but this trend
  is seen only for the SCC clusters (Figure~\ref{fig:Mbh_rlum}).  This
  trend is approximately given by $ \lr \propto \mbh^{\st 4.10 \pm
    0.42}$.
\item The NIR bulge luminosity of the BCG (closest to the X-ray peak
  emission) shows a correlation at an unprecedented level with the
  global cluster properties, such as X-ray mass shown in the left
  panel of Figure~\ref{fig:Mag}, and luminosity shown in the right
  panel of Figure~\ref{fig:Mag} (Equations~\ref{eqn8} and
  \ref{eqn10}).
\end{enumerate}

While we have provided strong evidence of there being an abundance of
CCRSs with enhanced radio luminosities in clusters where cooling
activity is at its full thrust (SCC clusters), a feedback
heating-cooling loop may require involvement of additional physical
processes, such as conduction as mentioned above, to halt the cooling
in WCC clusters.

\begin{acknowledgements} {The authors want to thank Heinz~Andernach
    for providing information on the initial BCG search positions for
    2MASS and Paul~Nulsen, Tod~Lauer and Douglas~Richstone for helpful
    discussions and the internal referee, Manuel~Perucho, for a
    beneficial feedback. The authors thank the external referee for a
    very interesting report and for provoking us to think about
    certain important issues in detail. R.~M. acknowledges support
    from the Deutsche Forschungsgemeinschaft~(DFG) through the
    Schwerpunkt Program 1177~(RE 1462/4). T.~H.~R and D.~S.~H.
    acknowledge support from the DFG through the Emmy Noether research
    grant RE 1462/2. Basic research in radio astronomy at the NRL is
    supported by 6.1 Base funding. This publication makes use of data
    products from the Two Micron All Sky Survey, which is a joint
    project of the University of Massachusetts and the Infrared
    Processing and Analysis Center/California Institute of Technology,
    funded by the National Aeronautics and Space Administration and
    the National Science Foundation. This research has made use of the
    NASA/IPAC Extragalactic Database (NED) which is operated by the
    Jet Propulsion Laboratory, California Institute of Technology,
    under contract with the National Aeronautics and Space
    Administration.  We acknowledge the usage of the HyperLeda
    database (http://leda.univ-lyon1.fr).  }
\end{acknowledgements}

\bibliographystyle{aa}
\bibliography{ref}


\input table.tex

\end{document}

%% file: makros.tex

\newcommand{\st}[1]{\mathrm{#1}} 
\newcommand{\bld}[1]{\textbf{#1}}
\newcommand{\pow}[2]{$\st{#1}^{#2}$}
\newcommand{\grad}{\hspace{-0.15em}\r{}}
\newcommand{\lm}{\lambda}

\newcommand{\average}[1]{\left\langle #1 \right\rangle}

\newcommand{\h}{~h_{71}~}
\newcommand{\hinv}[1]{~h_{71}^{#1}~}
\newcommand{\eq}[1]{(\ref{eq-#1})}
\newcommand{\wrt}{with respect to\ }
\newcommand{\mms}{\frac{M_{\odot}}{M}}
\newcommand{\mpy}{\st{M}_{\odot}~\st{yr}^{-1}}
\newcommand{\ct}{t_{\st{cool}}}
\newcommand{\rc}{r_{\st{cool}}}
\newcommand{\tvir}{T_{\st{vir}}}
\newcommand{\rvir}{R_{\st{500}}}
\newcommand{\mvir}{M_{\st{500}}}
\newcommand{\mbh}{M_{\st{BH}}}
\newcommand{\ms}{\st M_{\odot}}
\newcommand{\ls}{L_{\odot}}
\newcommand{\lxb}{L_{\st {Xb}}}
\newcommand{\lx}{L_{\st {X}}}
\newcommand{\lr}{L_{\st {R}}}
\newcommand{\lbcg}{L_{\st{BCG}}}
\newcommand{\Dl}{D_{\st{l}}}
\newcommand{\hiflux}{\textit{HIFLUGCS}}
\newcommand{\chandra}{\textit{Chandra}}
\newcommand{\vla}{\textit{VLA}}
\newcommand{\gmrt}{\textit{GMRT}}
\newcommand{\atca}{\textit{ATCA}}
\newcommand{\XMM}{\textit{XMM-Newton}}
\newcommand{\einstein}{\textit{Einstein}}
\newcommand{\asca}{\textit{ASCA}}
\newcommand{\rosat}{\textit{ROSAT}}
\newcommand{\mdr}{\dot{M}_{\st{classical}}}
\newcommand{\smdr}{\dot{M}_{\st{spec}}}

%% file: table.tex
\setlength\LTleft{-2.5in}
\setlength\LTright\fill
\begin{landscape}
\begin{longtable}{|l|c|l|l|l|c|c|c|c|c|}
    \caption{Radio data details for the central radio sources in the
      $\hiflux$ sample of galaxy clusters.}
    \label{tab:radio_data_ref}\\
    \hline \hline
    Cluster & Redshift & RA~(J2000) & Dec~(J2000) & $\nu$~(GHz) & $S_{\nu}$~(mJy) &
    $L_{\nu~(1.4~\textrm{GHz})}$ & $L_{\textrm{tot}}$ & S.I. & Ref.\\  
    \hline
    \endfirsthead
    \caption{continued.}\\
    \hline\hline
    Cluster & Redshift & RA~(J2000) & Dec~(J2000) & $\nu$~(GHz) & $S_{\nu}$~(mJy) &
    $L_{\nu~(1.4~\textrm{GHz})}$ & $L_{\textrm{tot}}$ & S.I. & Ref.\\  
    \hline
    \endhead
    \hline
    \endfoot
    A0085    & 0.055061 & 00$^h$41$^{'}$50.39$^{''}$ & -09$^{\circ}$18$^{'}$11.0$^{''}$  &  4.85    &    15.0$\pm$2.0        &   0.039800$\pm$0.001730  &   0.043300$_{ -0.011200}^{ +0.006280}$      &   -1.04	& 1        \\ 		  
             & 	& 00$^h$41$^{'}$50.52$^{''}$ & -09$^{\circ}$18$^{'}$10.9$^{''}$  &  1.4     &    56.7$\pm$2.5        &			      & 					 & 		& 2	 \\ 		    								     	   
             & 	& 00$^h$41$^{'}$50.48$^{''}$ & -09$^{\circ}$18$^{'}$11.1$^{''}$  &  0.074   &  1130.0$\pm$160.0      &			      & 					 & 		& 3	 \\ 		    								     	   
    \rul       	  	                                                                                                			        					   		  	   		    								     	   
               	  	                                                                                                			        					   		  	   		    								     	   
    A0119    & 0.044200 & 00$^h$56$^{'}$16.04$^{''}$ & -01$^{\circ}$15$^{'}$20.6$^{''}$  &  $-$     &    $-$                 &   0.000890               &   0.000952                               &   $-$	        & NO CRS	 \\ 	 
             & 	& 00$^h$56$^{'}$16.10$^{''}$ & -01$^{\circ}$15$^{'}$19.7$^{''}$  &          &                        &			      & 					 & 		& 	 \\ 		    								     	   
             & 	&        $-$                &            $-$                  &          &                        &   			      & 					 & 		& 	 \\
\hline \hline
 \end{longtable}
\end{landscape}